 \documentclass[twocolumn,twocolappendix]{aastex631}

\shorttitle{Light Curve Analysis of KT Eri}
\shortauthors{Hachisu et al.}

\begin{document}


\title{A multiwavelength light curve analysis of the classical nova KT Eri:\\
Optical contribution from a large irradiated accretion disk}


\author[0000-0002-0884-7404]{Izumi Hachisu}
\affil{Department of Earth Science and Astronomy,
College of Arts and Sciences,
University of Tokyo, Komaba 3-8-1, Meguro-ku, Tokyo 153-8902, Japan}
\email{izumi.hachisu@outlook.jp}

\author[0000-0002-8522-8033]{Mariko Kato}
\affil{Department of Astronomy, Keio University, 
Hiyoshi 4-1-1, Kouhoku-ku, Yokohama 223-8521, Japan}


\author[0000-0001-7796-1756]{Frederick M. Walter}
\affil{Department of Physics and Astronomy, Stony Brook University, Stony
Brook, New York, NY 11794, USA}


%

%

%



\begin{abstract}
KT Eri is a classical nova which went into outburst in 2009.
Recent photometric analysis in quiescence indicates a relatively longer
orbital period of 2.6 days, so that KT Eri could host a very bright accretion
disk during the outburst like in the recurrent nova U Sco,
the orbital period of which is 1.23 days.  We reproduced the optical $V$ 
light curve as well as the supersoft X-ray light curve of KT Eri in outburst,
assuming a large irradiated disk during a nova wind phase of the outburst
while a normal size disk after the nova winds stop.  This result is consistent
with the temporal variation of wide-band $V$ brightness that varies
almost with the intermediate-band Str\"omgren $y$ brightness,
because the $V$ flux is dominated by continuum radiation, the origin of
which is a photospheric emission from the very bright disk.
We obtained the white dwarf mass to be $M_{\rm WD}= 1.3\pm0.02 ~M_\sun$,
the hydrogen-burning turnoff epoch to be $\sim 240$ days
after the outburst, the distance modulus in the $V$ band to be 
$(m-M)_V=13.4\pm 0.2$, and the distance to KT Eri to be $d=4.2\pm0.4$ kpc
for the reddening of $E(B-V)= 0.08$.  The peak absolute $V$ brightness is
about $M_{V, \rm max}= -8.0$ and the corresponding recurrence time is 
$\sim 3,000$ yr from its ignition mass together with the mean mass-accretion 
rate of $\dot{M}_{\rm acc}\sim 1\times 10^{-9} ~M_\sun$ yr$^{-1}$
in quiescence.  Thus, we suggest that KT Eri is not a recurrent nova.
\end{abstract}






\begin{figure*}
\gridline{\fig{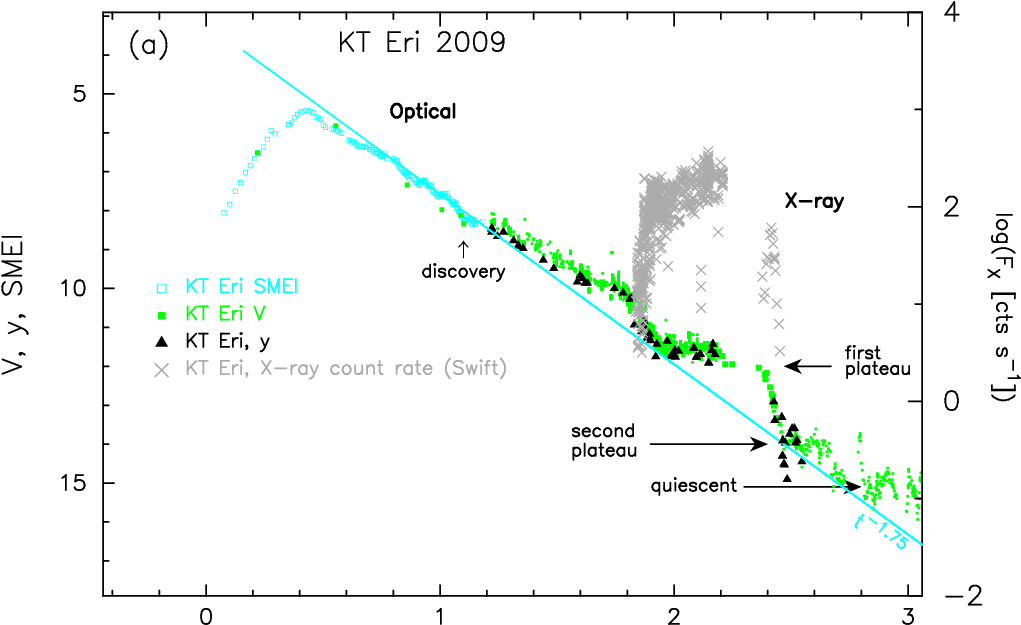}{0.7\textwidth}{}
          }
\gridline{\fig{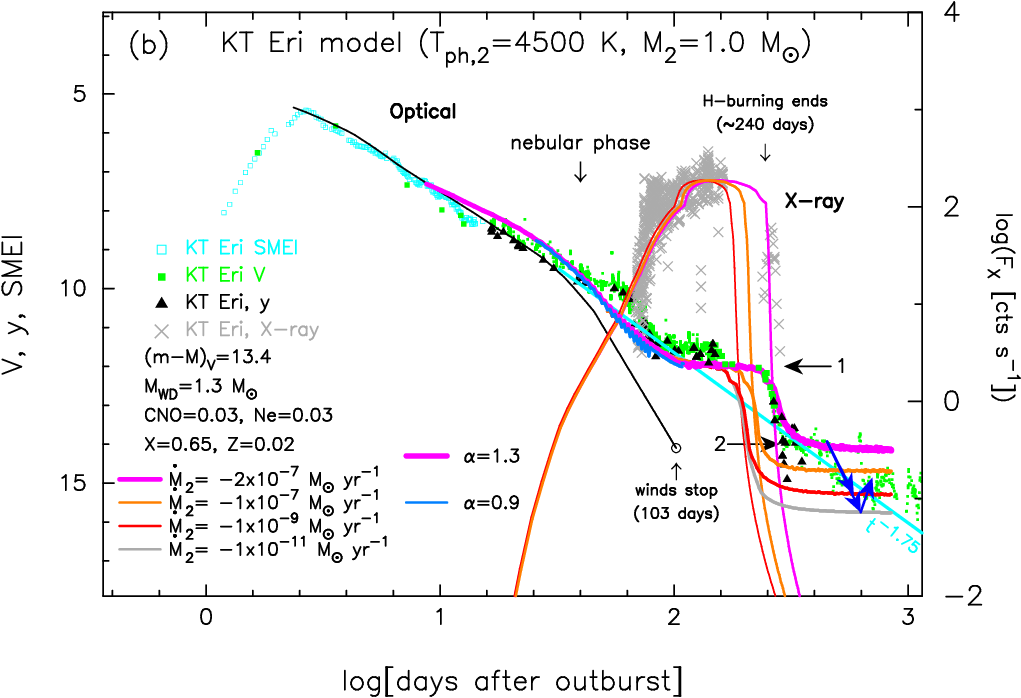}{0.7\textwidth}{}
          }
\caption{
(a) The $V$, $y$, and SMEI magnitudes of KT Eri against a logarithmic time.
We assume the outburst day of $t_{\rm OB}=$JD 2,455,147.5,
2.7 days before optical (SMEI) maximum.
The $V$ data are taken from VSOLJ, AAVSO, SMARTS \citep{wal12bt},
and IAU Circular No.9098 \citep{oot09ow}.  The $y$ data from \citet{ima12t}.
The SMEI data from \citet{hou10bh}.
The Swift X-ray (0.3--10.0 keV) count rates are also added,
taken from the Swift website \citep{eva09}.
The global decay trend is described by the universal
decline law of $L_V\propto t^{-1.75}$ \citep[straight cyan line
labeled $t^{-1.75}$:][]{hac06kb, hac23k}.
(b) Same as panel (a), but we overplot our model light curves 
(see Appendices \ref{opticall_thick_wind_model} and
\ref{large_irradiated_disk_binary} for details).
We assume a Roche-lobe-filling companion star with the photospheric
temperature of $T_{\rm ph,2}= 4,500$ K and mass of $M_2= 1.0 ~M_\sun$.
The black line is our free-free (FF) + photospheric blackbody (BB)
model light curve
of a $1.3 ~M_\sun$ white dwarf (WD). 
The thick magenta, orange, red, and gray lines denote
the model $V$ light curves for the $1.3 ~M_\sun$ WD with different
mass-accretion rates (see Table \ref{2nd_plateau_parameters} and
Appendix \ref{large_irradiated_disk_binary}) assuming a large
disk size of $\alpha=1.3$ (magenta line) during the nova wind phase.
We added another case of $\alpha=0.9$ (cyan-blue line) for comparison.
The thin colored lines
are for the corresponding model X-ray fluxes (0.3--10.0 keV).
The blue lines with an arrow depict temporal variations of the brightness
against temporal changes of mass-accretion rate in quiescence.
\label{kt_eri_only_v_x_big_disk_4500k_logscale}}
\end{figure*}

\section{Introduction}

KT Eri is a classical nova, discovered on UT 2009 November 25 (JD 2,455,160.5)
at $V=8.1$ by K. Itagaki \citep{yam09ig}.  Based on the data of Solar Mass
Ejection Imager (SMEI), \citet{hou10bh} constructed a pre-discovery light
curve (Figure \ref{kt_eri_only_v_x_big_disk_4500k_logscale}a).
The SMEI quantum efficiency has
a peak at $\lambda\sim 700$ nm with an FWHM of $\Delta \lambda = 300$ nm.
KT~Eri had already brightened up to 8.4 mag
on UT 2009 November 13.67 (JD 2,455,149.17) and then reached maximum
at 5.4 mag on UT 2009 November 14.67 (JD 2,455,150.17). 
The pre-discovery CCD magnitudes were also reported by \citet{oot09ow}. 
\citet{hou10bh} estimated the decay times from the peak
by 2 and 3 mag to be $t_2= 6.6$ days and $t_3=13.6$ days, respectively.
Thus, KT Eri belongs to the class of very fast novae.\footnote{The nova
speed class is defined by $t_2$ or $t_3$,
for example, very fast novae ($t_2 \le 10$ days),
fast novae ($10 < t_2 \le 25$ days),
moderately fast novae ($25 < t_2 \le 80$ days),
slow novae ($80 < t_2 \le 150$ days),
and very slow novae ($150 < t_2 \le 250$ days),
as defined by \citet{pay57}.}

Many photometric data were obtained after the discovery 
\citep[e.g.,][]{hun12cw}. We plot the $V$ data (filled green squares)
in Figure \ref{kt_eri_only_v_x_big_disk_4500k_logscale}a,
taken from the archives of the Variable Star Observers League
of Japan (VSOLJ), American Association of Variable Star Observers
(AAVSO), SMARTS \citep{wal12bt}, those reported in IAU Circular No.9098
\citep{oot09ow}, and \citet{ima12t}.
\citet{ima12t} reported their $BVyR_{\rm C}$ photometry.
We also plot their $y$ magnitude data (filled black triangles)
in Figure \ref{kt_eri_only_v_x_big_disk_4500k_logscale}a.

Optical spectroscopy of KT Eri were reported by \citet{rag09bs},
\citet{bod10op},
\citet{ima12t}, \citet{ara13ii}, \citet{rib13bd}, and \citet{mun14mv}.
The broad Balmer emission showed FWHM$\sim 3200$--$3400$ km s$^{-1}$
\citep{mae09ai, mae09i} or the \ion{He}{1} 1.083$\mu$m line had 
FWHM$\sim 4000$ km s$^{-1}$ and its P-Cygni absorption
profile was extending to 3600 km s$^{-1}$ on UT 2009 November 26.5
\citep[JD 2,455,162; ][]{rud09pr}.
\citet{rib13bd} analyzed the H$\alpha$ line profiles between
42 and 73 days after optical maximum  and proposed 
a model of dumbbell shaped expanding ejecta with the expanding
velocity of $v_{\rm exp} = 2800\pm 200$ km s$^{-1}$ and inclination
angle of $58^{+6}_{-7}$ deg.

X-rays were observed with Swift \citep{bod10op, bea10bo, schw11},
Chandra \citep{nes11ds, sun20od, pei21on},
and XMM-Newton \citep{sun20od}.  The bright supersoft X-ray source
(SSS) phase started $\sim 65$ days and ended $\sim 280$ days
after optical maximum (Figure \ref{kt_eri_only_v_x_big_disk_4500k_logscale}a).
KT Eri was also observed with radio 
\citep{obr10ms} and near infra-red \citep{rud09pr, raj13ba}. 

\citet{jur12rd} searched the archival plates of the Harvard College
Observatory for a previous outburst of KT Eri. 
They found no outbursts
between 1888 and 1962, and concluded that, if KT Eri
is a recurrent nova, it should have a recurrence time of centuries.
\citet{dar12rb} discussed the positions of KT Eri in various color-magnitude
diagrams and concluded that the companion star of KT Eri is a red giant,
less evolved than that of RS Oph and T CrB.

\citet{schaefer22wh} pointed out two distinct plateau phases
in the $BVI_{\rm C}$ light curves: the first one is coincident with
the supersoft X-ray source (SSS) phase 
(Figure \ref{kt_eri_only_v_x_big_disk_4500k_logscale}a).
They obtained the orbital period to be $P_{\rm orb}=2.6$ days both
from the $\sim 0.2$ mag TESS photometric variation and
the orbital velocity variation.  This $P_{\rm orb}$
is exceptionally long for classical novae.  Thus,
KT Eri resembles recurrent novae both in the optical plateau during
the SSS phase and in the long orbital period.  
\citet{schaefer22wh} also pointed out a large temporal variation of
the $V$ light curve in quiescence, as large as $\Delta V\sim 1.5$,
and raised the question on the origin of such a large variation,
the timescale of which is from 1 month to 3 months.

\citet{schaefer22wh} estimated the mass-accretion rate on to the white
dwarf (WD) to be $\sim 3.5\times 10^{-7} ~(d/5.1{\rm ~kpc})^2 
~M_\sun$ yr$^{-1}$ from the brightness of a viscous disk and companion star,
and concluded that KT Eri is a recurrent nova with recurrence time of
40--50 yr.  Here, $d$ is the distance to KT Eri.
From the theoretical point of view, however, a nova outburst never occurs
for such a high mass-accretion rate of $\sim 3.5\times 10^{-7} ~M_\sun$
yr$^{-1}$ \citep[e.g.,][for recent estimates]{wol13bb, kat14shn}.
\citet{sun20od} reported a much smaller mass-accretion rate of
$\dot{M}_{\rm acc} \approx 1.9 \times 10^{-10} 
~(d/3.7{\rm ~kpc})^2 ~M_\sun$ yr$^{-1}$
based on their XMM Newton X-ray observation on UT 2018 February 15,
about 8.3 yr after the outburst.

In the present paper, we propose a sophisticated model for the KT Eri 2009
outburst that explains the puzzling observational properties mentioned above. 
Our model consists of a WD, accretion disk, and companion star.
We include the irradiation effects of the disk and companion star, while
\citet{schaefer22wh} did not take into account the irradiation effects 
in their analysis.

Our paper is organized as follows.  First we present a summary of 
optical and X-ray observations with our quick interpretation 
in Section \ref{observation_interpretation_kt_eri}.
Discussion and conclusions follow in 
Sections \ref{discussion} and \ref{conclusions}, respectively.
Next, in Appendix \ref{opticall_thick_wind_model},
we describe our free-free emission
model light curves for the $V$ and $y$ bands and reproduce an early phase
of the $V,y$ light curves and later supersoft X-ray light curve.
Appendix \ref{large_irradiated_disk_binary} is devoted to calculation
of the model $V,y$ light curves based on our irradiated disk model,
to reproduce the $V,y$ light curves of KT Eri in the SSS and
later phases.  To estimate the distance to the nova,
the time-stretching method is applied to KT Eri in Appendix 
\ref{distance_moduli_bvi}.

\begin{figure*}
\epsscale{1.15}
\plottwo{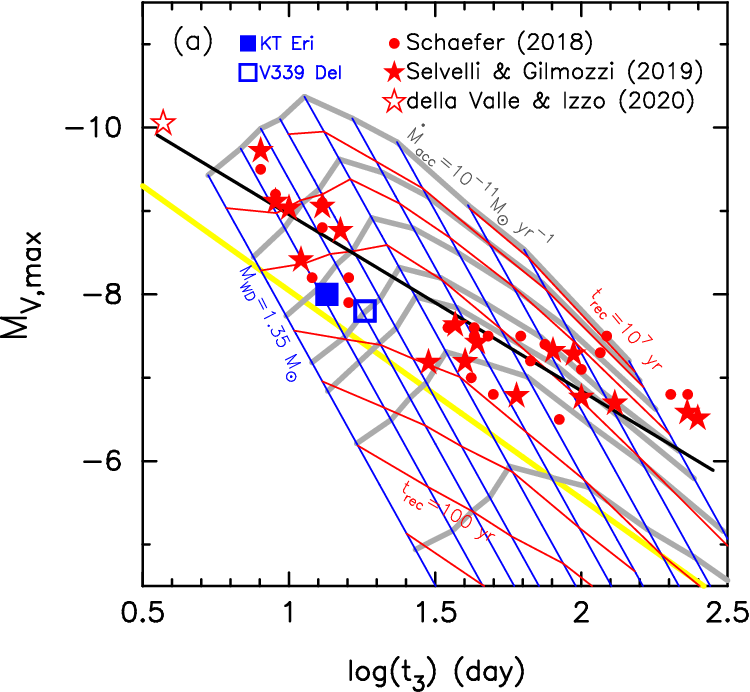}{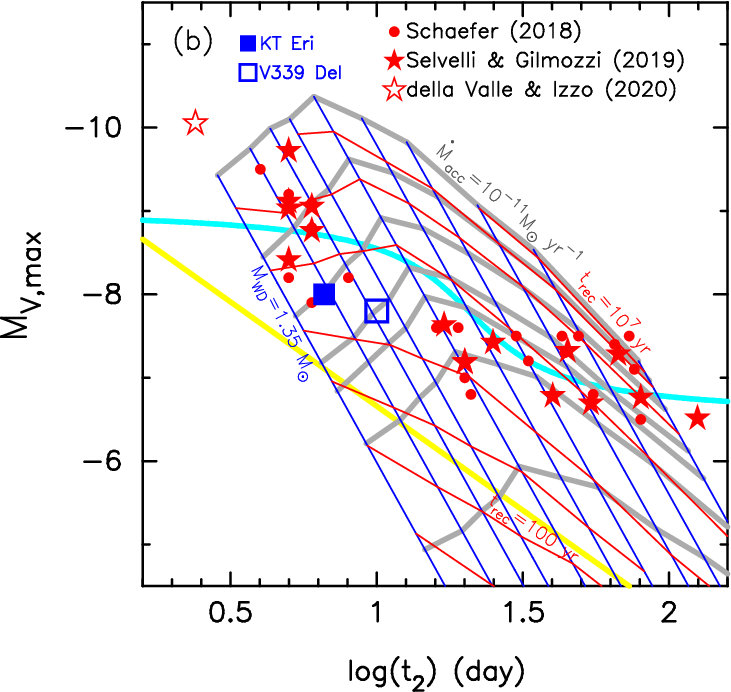}
\caption{
The maximum absolute $V$ magnitude against the rate of decline (MMRD)
diagrams for selected novae; 
(a) $t_3$-$M_{V, \rm max}$ and (b) $t_2$-$M_{V,\rm max}$.
The filled blue square represents the position of KT~Eri.
The open blue square is V339 Del, taken from \citet{hac24km}.
We add the MMRD points of some selected galactic novae.
The filled red circles represent novae taken from
``Golden sample'' of \citet{schaefer18}, from which we exclude recurrent
novae and V1330~Cyg.  The filled stars are novae taken from
\citet{sel19}.  The open star is V1500~Cyg taken from \citet{del20}.
The blue lines indicate model equi-WD mass lines, from left to right,
1.35, 1.3, 1.25, 1.2, 1.1, 1.0, 0.9, 0.8, 0.7, and $0.6~M_\sun$;
the thick solid gray lines denote model equi-mass accretion rate
($\dot M_{\rm acc}$) lines, from lower to upper, $3\times 10^{-8}$,
$1\times 10^{-8}$, $5\times 10^{-9}$, $3\times 10^{-9}$, $1\times 10^{-9}$,
$1\times 10^{-10}$, and $1\times 10^{-11} M_\sun$~yr$^{-1}$;
the red lines represent model equi-recurrence time lines, from lower to upper,
$t_{\rm rec}= 30$, 100, 300, 1000, 10000, $10^5$, $10^6$, and $10^7$~yr.
These lines are taken from \citet{hac20skhs} based on the universal
decline law of novae and model calculation of mass accretion onto each WD.
The thick yellow line corresponds to the $x_0=2$ line \citep{hac20skhs},
below which the models are not valid.
Two empirical MMRD relations are added by the thick solid black line
\citep{sel19} in panel (a) and by the thick solid cyan line
\citep{del20} in panel (b). 
Many recurrent novae are located in the lower-left portion of panels (a)
and (b)  \citep[see, e.g.,][]{shafter23ch}, and are below the solid yellow
lines.  Therefore, \citet{hac24km}'s model lines could not be applied to
recurrent novae.
\label{vmax_t3_vmax_t2_selvelli2019_schaefer2018_2fig}}
\end{figure*}

\section{Observational summary and quick interpretation}
\label{observation_interpretation_kt_eri}

Figure \ref{kt_eri_only_v_x_big_disk_4500k_logscale}a shows
the optical and X-ray light curves for the KT Eri 2009 outburst
in a logarithmic time.  
The origin of the time is the outburst day.
Here, we adopt the outburst day of $t_{\rm OB}=$JD 2,455,147.5 
(UT 2009 November 12.0), i.e., 
2.7 days before the SMEI maximum \citep{hou10bh}. 

We estimated the outburst day from the rising trend in the SMEI light curve
by extending the rising trend down to the quiescent brightness of
$V\sim 15$ mag.
If an X-ray flash of KT Eri was observed like in the classical nova YZ Ret
\citep{kon22wa}, we are able to constrain the outburst day much more
accurately \citep[e.g.,][]{kat22shb, kat22shc}.
This $t_{\rm peak}= 2.7$ days, the days from the outburst to optical
maximum, is broadly consistent with $t_{\rm peak}= 2.5$ days for
a self-consistent nova model of a $1.3 ~M_\sun$ WD
with the mass-accretion rate of $\dot M_{\rm acc}= 2\times 10^{-9} ~M_\sun$
yr$^{-1}$ \citep[model M13C10; ][]{kat24sh}.

The characteristic properties of the light curves of KT Eri as follows:
\begin{itemize}
\item[(1)] The SMEI magnitude reaches maximum 2.7 days after
the outburst and then declines almost along the universal decline law of
$L_{\rm SMEI} \propto t^{-1.75}$ with $t_2= 6.6$ days and $t_3=13.6$ days
\citep{hou10bh}.
\item[(2)] The optical $V$ and Str\"omgren $y$ magnitudes of KT Eri declined
almost in the same way and did not depart from each other \citep{ima12t}.
\item[(3)] The $V,y$ light curves show two distinct plateau
phases, the first is at $V\sim 12.0$ between days 80 and 240
after maximum and the second is at $V\sim 14.1$ from 300 to 440 days
after maximum \citep{schaefer22wh}.
\item[(4)] The first plateau appears almost during the same period
as the SSS phase.
However, a close look at the timing indicates that it starts slightly
later than the rise of soft X-ray count rate
and ends slightly before the decay of X-ray.
\item[(5)] After the second plateau, the $V$ brightness drops to $V\sim 15.1$,
accompanying a large amplitude variation of $\Delta V \sim 1.5$--2 
and timescales of up and down from 1 month to 3 months \citep{schaefer22wh}.
\end{itemize}
In what follows, we explain how these properties make KT Eri an outlier 
of classical novae and give a quick understanding of our 
binary model.

\subsection{Distance and reddening}
\label{distance_reddening}

The distance and reddening are critically important to estimate the
absolute brightness of the nova. 
The distance was estimated to be
$d=4.1^{+0.5}_{-0.4}$ kpc by \citet{bai21rf} based on the Gaia eDR3
parallax.  The extinction $E(B-V)=0.08$ was given by \citet{rag09bs}
from the \ion{Na}{1} D1 line width and the relation of \citet{mun97z}.
Then, we have the distance modulus
in the $V$ band of $(m-M)_V=13.3 \pm 0.3$ from
\begin{equation}
(m-M)_V= 5 \log (d / {\rm 10~pc}) + 3.1 E(B-V),
\label{distance_modulus_v_band}
\end{equation}
where $d$ is the distance and the coefficient of $R_V=3.1$ is taken from
\citet{rie85}.

Assuming a different prior from that of the Gaia team in a Bayesian
calculation, \citet{schaefer22wh} obtained $d= 5110^{+920}_{-430}$ pc
based on the same eDR3 parallax.  This gives a larger distance modulus
in the $V$ band of $(m-M)_V= 13.8^{+0.4}_{-0.2}$.
It is however noted that \citet{schaefer22b} adopted $d=4211^{+466}_{-296}$
pc for KT Eri.  If we adopt this distance, we have $(m-M)_V=13.4\pm 0.2$.

We also estimate the distance modulus to KT Eri by comparing the 
intrinsic brightnesses of well studied novae, LV Vul and V339 Del, whose 
distance moduli are well determined by \citet{hac24km}.
We obtained a similar distance modulus in the $V$ band to KT Eri to be
$(m-M)_V= 13.4\pm 0.2$ (see Appendix \ref{distance_moduli_bvi}).

Figure \ref{kt_eri_only_v_x_big_disk_4500k_logscale}b shows our 
$1.3 ~M_\sun$ WD model that is chosen to fit with the early light curve, 
adopting $(m-M)_V=13.4$ (see Section \ref{opticall_thick_wind_model}
for details).  If we adopt $(m-M)_V=13.8$, our model $V$ light curve
(black line) will be parallel but 0.4 mag below the observation.
Thus, we adopted $d=4.2\pm 0.4$ kpc, $E(B-V)=0.08$, and $(m-M)_V=13.4\pm 0.2$.


\begin{figure*}
\gridline{\fig{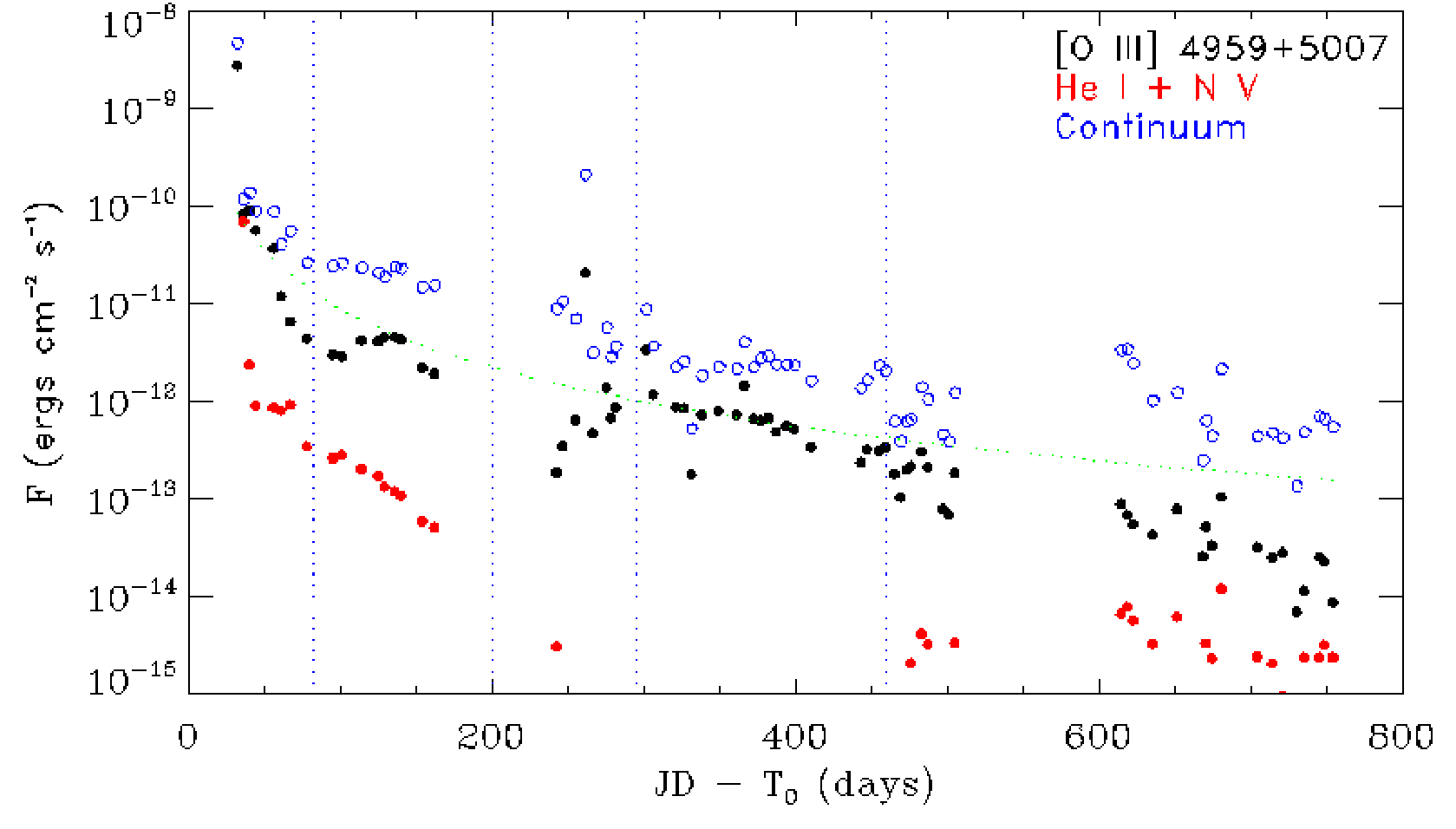}{0.85\textwidth}{(a) Continuum (blue),
[\ion{O}{3}] (black), and \ion{He}{1}+\ion{N}{5} (red) fluxes in the $V$ band}
          }
\gridline{\fig{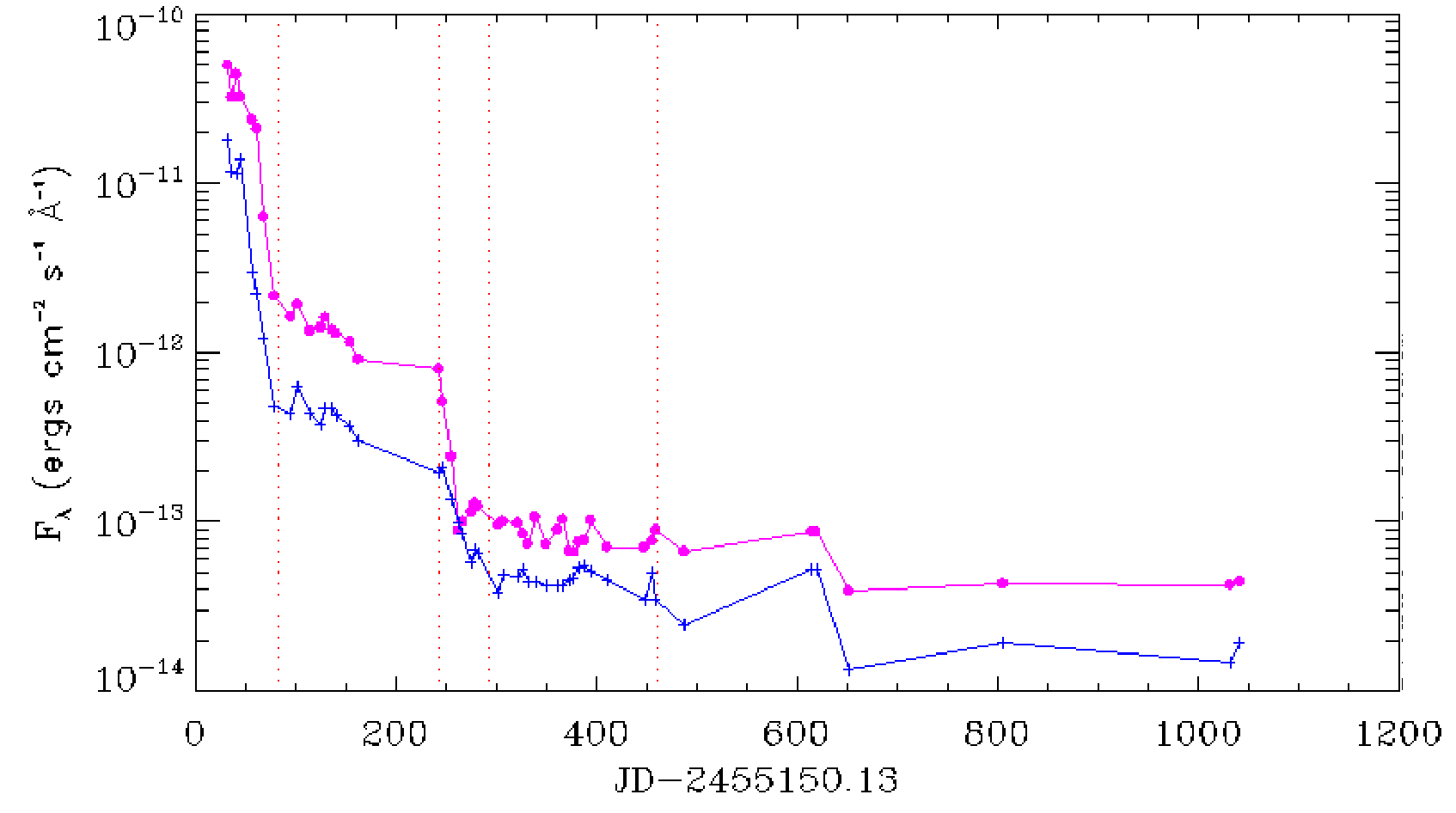}{0.85\textwidth}{(b) Intensities of 
\ion{He}{2} line (magenta) and continuum (blue)}
          }
\caption{
The fluxes of emission lines and continuum of KT Eri calculated
from the SMARTS spectra \citep{wal12bt}.  The first two vertical dotted
lines represent the period of the first plateau while the second two
vertical dotted lines do that of the second plateau.
(a) Temporal flux variations of continuum (blue),
[\ion{O}{3}] 4959\AA\  $+$ 5007\AA\  lines (black),
and \ion{He}{1} 4922\AA\ $+$ \ion{N}{5} 4945\AA\  lines (red)
in the $V$ band \citep[see, e.g., ][for emission lines of novae]{wil12}.
The continuum flux dominates the line fluxes at every time.
$T_0$ is the time at maximum brightness (JD 2,455,150.13).
(b) Temporal intensity variations of continuum (blue) and
\ion{He}{2} 4686\AA\  line (magenta).
\label{vflux_o3_he2}}
\end{figure*}

\subsection{Peak magnitude: WD mass, mass-accretion rate, and recurrence time}
\label{WD_mass_accreion_rate}

Adopting $(m-M)_V=13.4$, the absolute peak $V$ brightness of KT Eri is 
calculated from $M_{V, \rm max}= m_{V,\rm max} - (m-M)_V = 5.4 - 13.4 = -8.0$.
\citet{hou10bh} obtained the decay times of $t_2= 6.6$ and $t_3= 13.6$ days. 
We plot these values in the maximum magnitude versus rate of decline
(MMRD) diagram with other classical novae in Figure 
\ref{vmax_t3_vmax_t2_selvelli2019_schaefer2018_2fig}.  
KT Eri (filled blue square) is located close to the classical nova V339 Del
(open blue square) and among the other well observed classical novae. 
Note that the distribution of novae in these
MMRD plots is affected by the observational selection effect, especially
in the lower-left portion of each plots \citep[e.g.,][]{shafter23ch},
because faint and fast decay novae are easily missed.

\citet{hac20skhs} obtained theoretical peak absolute $V$ magnitudes 
$M_{V, \rm max}$ and corresponding theoretical $t_3$-$M_{V,\rm max}$ and 
$t_2$-$M_{V,\rm max}$ based on a number of theoretical light curves
for different mass-accretion rates $\dot{M}_{\rm acc}$ and WD masses
$M_{\rm WD}$. 
Figure \ref{vmax_t3_vmax_t2_selvelli2019_schaefer2018_2fig} also shows 
theoretical lines that connect a specific WD mass, recurrence period, 
or mass accretion rate, from their theoretical database 
\citep{hac20skhs}.

If we fix the WD mass, a nova becomes brighter (smaller $M_{V,\rm max}$)
and faster (smaller $t_3$ and $t_2$) for a smaller mass-accretion rate
(and a longer recurrence time). 
The position of KT Eri (filled blue square)
is located close to the blue line
of $M_{\rm WD}= 1.3 ~M_\sun$ and the gray line of 
$\dot M_{\rm acc}= 1\times 10^{-9} ~M_\sun$ yr$^{-1}$.
The recurrence time is estimated 
from two red lines of $\tau_{\rm rec}= 1000$ yr and 
$\tau_{\rm rec}= 10,000$ yr 
to be about $\tau_{\rm rec}\sim 3,000$ yr. 
Both the diagrams of Figure
\ref{vmax_t3_vmax_t2_selvelli2019_schaefer2018_2fig}a 
($t_3$-$M_{V, \rm max}$) and Figure 
\ref{vmax_t3_vmax_t2_selvelli2019_schaefer2018_2fig}b 
($t_2$-$M_{V, \rm max}$) give a similar result.  
It should be again noted that the $t_{\rm peak}= 2.7$ days (from the outburst
to optical maximum) is roughly consistent with
a self-consistent nova model of a $1.3 ~M_\sun$ WD
with the mass-accretion rate of $\dot M_{\rm acc}= 2\times 10^{-9} ~M_\sun$
yr$^{-1}$ \citep[model M13C10; $t_{\rm peak}= 2.5$ days; ][]{kat24sh}. 
Thus, it is unlikely that KT Eri is a recurrent nova with recurrence
time of $40-50$ yr \citep{schaefer22wh}.  
This will be discussed in Section \ref{kt_eri_recurrent_nova_no}. 


\subsection{$V,y$ light curves in the early phase: WD mass}
\label{Vy_light_curve_early}

A nova starts from unstable hydrogen burning on a WD.  Then, a
hydrogen-rich envelope of the WD expands and emits strong winds
\citep[e.g.,][for a recent nova calculation]{kat22sha}.
Free-free emission from the nova winds dominates the optical flux of a nova
\citep[e.g.,][]{gal76, enn77}.  \citet{hac06kb} modeled nova light curves
for the free-free emission based on the optically thick winds calculated
by \citet{kat94h}, the $V$ flux of which can be simplified as
\begin{equation}
L_{V, \rm ff,wind} = A_{\rm ff} ~{{\dot M^2_{\rm wind}}
\over{v^2_{\rm ph} R_{\rm ph}}}.
\label{free-free_flux_v-band}
\end{equation}
This $V$ flux represents the flux of free-free emission from optically thin
plasma just outside the photosphere, and $\dot{M}_{\rm wind}$ is the
wind mass-loss rate, $v_{\rm ph}$ the velocity at the photosphere,
and $R_{\rm ph}$ the photospheric radius.  See \citet{hac20skhs} for
the derivation of this formula and the coefficient $A_{\rm ff}$.
In our $V$ light curve model, the total $V$ band flux is defined by 
the summation of the free-free (FF) emission luminosity
and the $V$ band flux of the photospheric luminosity $L_{\rm ph, WD}$
(assuming blackbody (BB)), i.e., FF+BB,
\begin{equation}
L_{V, \rm total} = L_{V, \rm ff,wind} + L_{V, \rm ph, WD}.
\label{luminosity_summation_flux_v-band}
\end{equation}
The photospheric $V$ band luminosity of the WD
is calculated from a blackbody with
$T_{\rm ph}$ and $L_{\rm ph}$ using a canonical response function
of the $V$ band filter, where $T_{\rm ph}$ and $L_{\rm ph}$ are
the photospheric temperature and luminosity, respectively.

The winds are called ``optically thick winds,''
because matter is accelerated deep inside the photosphere
\citep{bat78, rug79b, kat94h}.  Note that
the wind itself becomes optically thin outside the photosphere.
Free-free emission comes from optically-thin plasma outside the photosphere.
Therefore, this free-free luminosity is not limited by the Eddington
luminosity because the Eddington luminosity can be applied only to
the optically-thick region.

We have estimated the WD mass by directly comparing the model light curves
\citep[e.g., ][]{hac15k, hac16k} with the observation.
See Appendix \ref{opticall_thick_wind_model} 
for details of this light curve model. 
Figure \ref{kt_eri_only_v_x_big_disk_4500k_logscale}a shows
the SMEI, $V$, and $y$ light curves that show almost the same decline 
toward the quiescent phase.
By comparing the KT Eri $V,y$ light curves with a large number of model
light curves in \citet{hac06kb, hac15k, hac16k, hac18kb} and 
\citet{hac20skhs},  we chose a best-fit model, which is plotted in
Figure \ref{kt_eri_only_v_x_big_disk_4500k_logscale}b (black line: 
$1.3 ~M_\sun$ WD model).  Its WD mass is $1.3 ~M_\sun$, being consistent
with the WD mass estimated from the position of KT Eri in the MMRD
diagram (Figure \ref{vmax_t3_vmax_t2_selvelli2019_schaefer2018_2fig}).
When free-free emission dominates the spectrum, both the $V$ and $y$ 
light curves show essentially the same decline along with our model
light curve, until day $\sim 40$, as in Figure 
\ref{kt_eri_only_v_x_big_disk_4500k_logscale}b.

Our basic nova evolution is governed by 
a time-evolutionary sequence of the decreasing envelope mass of
\begin{equation}
{{d} \over {d t}} M_{\rm env} = \dot M_{\rm acc}
- \dot M_{\rm wind} - \dot M_{\rm nuc},
\label{nova_evoluion_eq}
\end{equation}
where $M_{\rm env}$ is the mass of a hydrogen-rich envelope on the WD, 
$\dot M_{\rm acc}$ the mass accretion rate onto the WD,
$\dot M_{\rm nuc}$ the mass decreasing rate of hydrogen-rich envelope
by nuclear (hydrogen) burning, and usually $\dot M_{\rm acc} \ll 
\dot M_{\rm wind}$, and $\dot M_{\rm acc} \ll \dot M_{\rm nuc}$ for
typical classical novae \citep[see, e.g.,][]{hac06kb}.
Therefore, the envelope mass is decreased by winds and nuclear burning.
A large amount of the envelope mass is lost mainly by winds because
of $\dot M_{\rm wind} \gg \dot M_{\rm nuc}$ in the early phase of
nova outbursts (see Equation (7) of \citet{hac06kb} for details).
These FF+BB model light curves broadly follow the $V,y$ light curves
for the distance modulus in the $V$ band of $(m-M)_V=13.4$
until day $\sim 30$--40, where the nebular phase started.


\begin{figure*}
\gridline{\fig{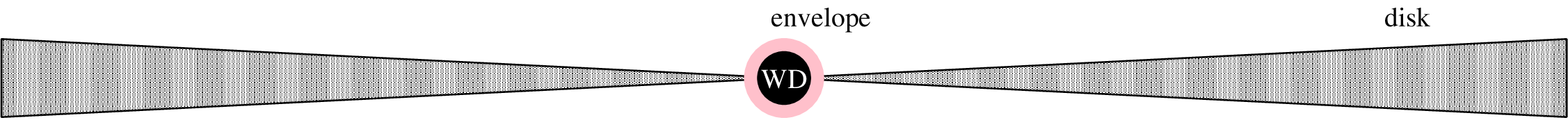}{0.95\textwidth}{(a) before ignition}
          }
\gridline{\fig{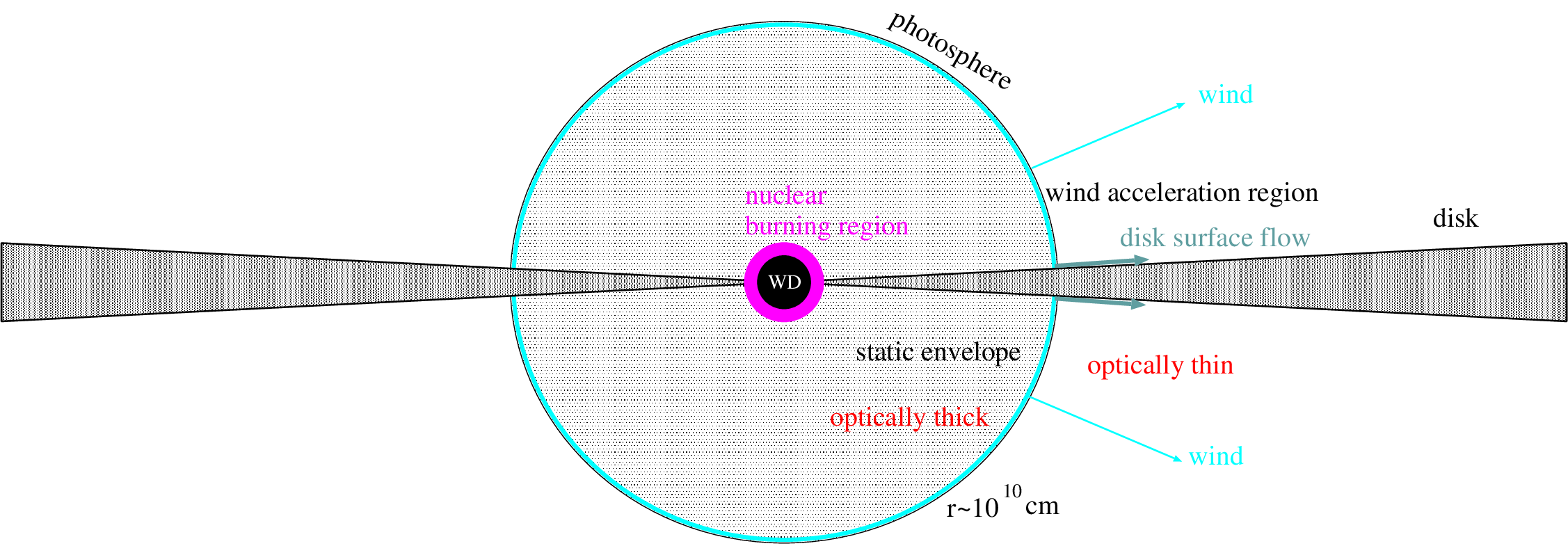}{0.95\textwidth}{(b) at the epoch when winds begin
to emerge from the photosphere}
          }
\gridline{\fig{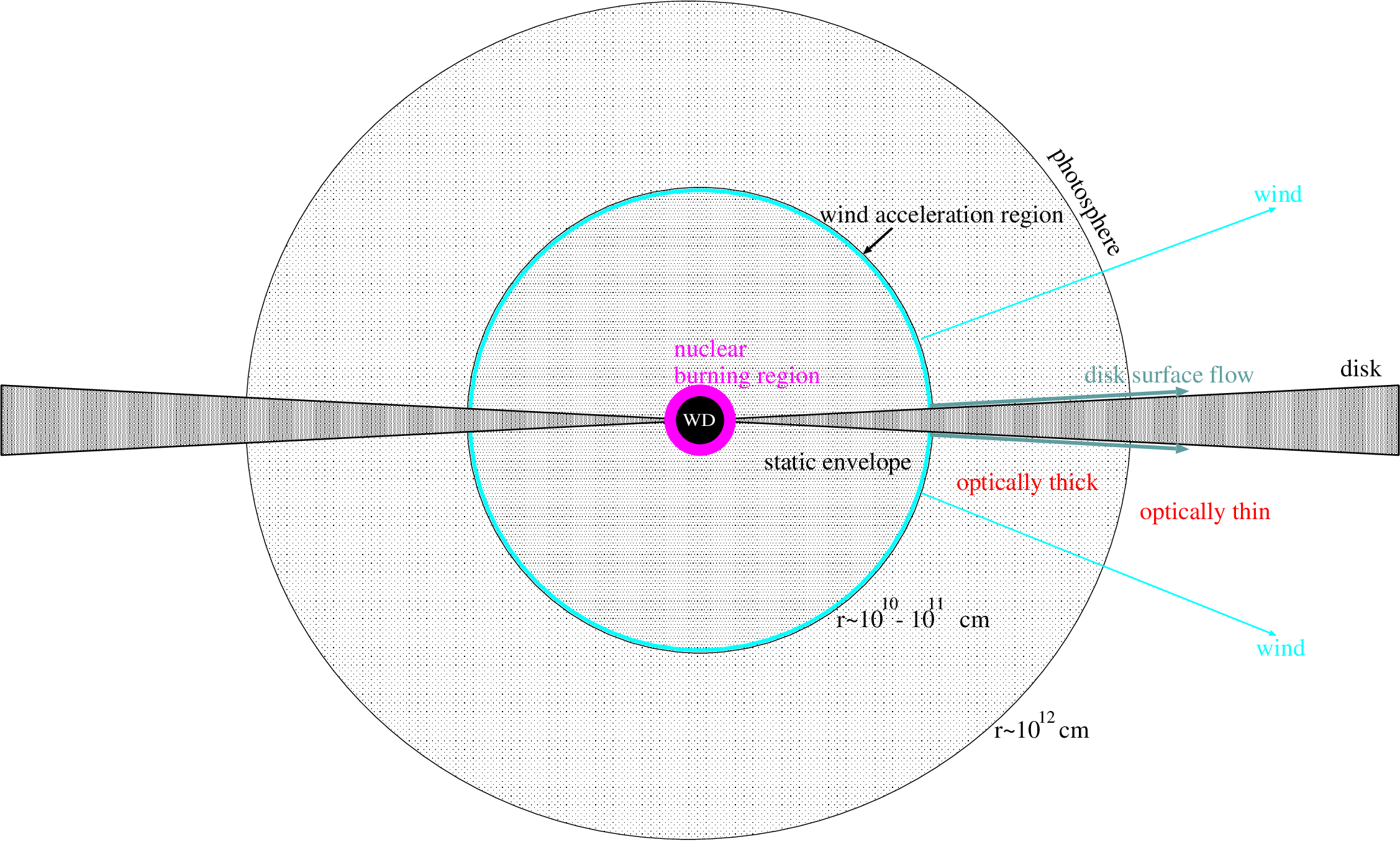}{0.95\textwidth}{(c) at maximum expansion of the
photosphere}
          }
\caption{
Schematic configurations of a WD envelope and accretion disk
during a nova outburst:
(a) before ignition of hydrogen burning; (b) at the epoch when winds begin to
emerge from the photosphere; (c) at the maximum expansion of the photosphere.
Optically thick winds have an opening angle avoiding the accretion disk.  The
spherical nova envelope configurations are taken from the spherically
symmetric $1.3 ~M_\sun$ WD model in Figure 6 of \citet{kat24sh}.
The thick light-cyan region describes the acceleration region of winds.
The dark-cyan arrows represent the disk surface (boundary layer) flows,
which exist near the surface of accretion disk
and have slower outflow velocities than those of winds. 
\label{kt_eri_disk_config}}
\end{figure*}

\subsection{Same decline trend in the $V$ and $y$ light curves}
\label{Vy_light_curve_same}

In Figure \ref{kt_eri_only_v_x_big_disk_4500k_logscale}a,
the $V$ and $y$ light curves similarly decline and never depart from
each other.  This is a remarkable property of KT Eri,
because many novae show large brightness difference between $V$ and $y$ 
magnitudes in the nebular phase, in which
strong emission lines such as [\ion{O}{3}] 4959, 5007 \AA\  contribute 
to the wide $V$ band flux. On the other hand, the intermediate Str\"omgren
$y$ band is designed to avoid such strong emission lines \citep[see
e.g., Figure 1 of ][]{mun13dcvf}.  Thus, 
the $V$ magnitude is brighter than the $y$ magnitude in the nebular phase,
(see, e.g., \citet{loc76m} for V1500 Cyg, \citet{gal80ko} for V1668 Cyg,
and \citet{mun15mm} and \citet{hac24km} for V339 Del).

Figure \ref{vflux_o3_he2}a shows temporal flux variations of
the continuum (blue open circles),
[\ion{O}{3}] 4959, 5007 \AA\  (black dots),
and \ion{He}{1} 4922\AA\ +\ion{N}{5} 4945\AA\  (red dots)
of KT Eri in the $V$ band, all of which are calculated
from the SMARTS spectra of KT Eri \citep{wal12bt}.
Here, we confirm that, in the $V$ band of KT Eri, the fluxes of emission
lines such as [\ion{O}{3}] are not larger than the continuum flux.
The continuum flux in the $V$ band always dominates over each line flux
in the $V$ band.
This supports our explanation that the $V$ and $y$ magnitudes do not
depart from each other.

Although we see no clear indication of nebular phase in the light curve, 
there is a sign in the color-magnitude diagram of the KT Eri 2009 outburst
(see Figure \ref{hr_diagram_kt_eri_v339_del_bv_mfs}b).
The start of the nebular phase can be detected as a turning point of
the track, that is, from the bluest peak of the color index $B-V$ 
\citep[e.g.,][for such an example]{hac21k}.
The minimum (bluest) $B-V$ color was observed on day $39$ \citep{ima12t}.  
We regard the starting time of the nebular phase as day $\sim 40$,
which is indicated by the downward arrow labeled ``nebular phase''
in Figure \ref{kt_eri_only_v_x_big_disk_4500k_logscale}b.
The definition of color $B-V$ turning point is not the same as the
definition of the start of nebular phase (e.g., [\ion{O}{3}]$/$H$\beta>1$),
but the both dates are broadly coincident with each other in several novae,
as later explained in Appendix \ref{color-magnitude_diagram}.

\citet{hac22k} theoretically showed that a strong shock inevitably
arises far outside the photosphere in the post-maximum phase of a nova.
In \citet{kat22sha}'s nova model, the photospheric wind velocity decreases
with time before optical maximum, but turns to increase after the
optical maximum, so that the wind ejected later catches up with
the wind ejected earlier and drives a shock.
Thus, a strong shock naturally arises
far outside the photosphere after the optical maximum.
The mass of the shocked shell increases with time and reaches about 90\%
of the total ejecta mass.  In other words, a large part of nova ejecta is
eventually confined to the shocked shell.  Therefore, the shocked shell
becomes a main source of strong emission lines \citep[e.g.,][]{hac24km}. 

In KT Eri, however, these emission lines are not so strong compared with
the continuum flux, because the continuum flux is rather high as depicted
in Figure \ref{vflux_o3_he2}a.  The flux of the continuum is always larger
than those of strong emission lines such as [\ion{O}{3}] and 
\ion{He}{1}+\ion{N}{5} in the $V$ band.
We will clarify the reason why the continuum flux always dominates 
each emission line flux later in this subsection and in Section
\ref{Vy_light_curve_mid_late}.

Figure \ref{kt_eri_only_v_x_big_disk_4500k_logscale}b shows that 
our model light curve (black line: FF+BB$=$continuum flux of free-free
emission plus photospheric blackbody emission in Equation
(\ref{luminosity_summation_flux_v-band}))
decays more rapidly than the $V,y$ light curves after day $\sim 40$.
This is because, in our model, the wind mass-loss rate $\dot{M}_{\rm wind}$
rapidly drops after day $\sim 40$.
Thus, there should be additional optical sources, which are not line emissions 
because the $y$ light curve does not depart from the $V$ light curve.
This continuum emission source should be optically-thick. 
We regard that this additional source is 
a photospheric emission of an equatorial, optically-thick disk. 

The optically-thick winds are accelerated by the radiation pressure gradient
deep inside the photosphere. Figure \ref{kt_eri_disk_config}
shows this ``acceleration region of winds.''  The thick light-cyan
circular lines in the figure correspond to this acceleration
region of winds.  The envelope is almost static (very low velocity)
inside the acceleration region \citep{kat22sha, kat24sh}.
Therefore, we may conclude that 
the accretion disk is almost intact inside the acceleration region.
The radius of the acceleration region is about $r_{\rm accel}\sim 0.1 ~R_\sun
\sim 10^{10}$ cm when winds start to emerge from the photosphere but expands
up to $r_{\rm accel}\sim 1 ~R_\sun \sim 10^{11}$ cm at maximum expansion
of the photosphere (Figure \ref{kt_eri_disk_config}c).
The momentum of wind is spherically radial so that it does not so push
the accretion disk outward because it has already an open angle avoiding
the disk inside the acceleration region.  The very thin surface layer 
can be blown in the wind (disk surface flow) and the disk expands over
the Roche lobe size \citep{hac03ka, hac03kb, hac03kc}.
Thus, we suppose that the accretion disk is almost intact, at least,
from the view point of spherically radial momentum. 
However, it should be noted that the disk is embedded in radiation
field and could be influenced in a thermal timescale.

We suppose that the interaction between two fluids (nova wind and
accretion disk) causes a Kelvin-Helmholtz instability and effectively
add radial momentum to a surface layer of the accretion disk.
This can accelerate a disk surface flow up to $\sim 1000$ km s$^{-1}$,
because the nova winds have velocities of $\sim 2000-3000$ km s$^{-1}$.

We assume that the disk is an accretion disk.  This means that the mass
is supplied by the companion star because the nova wind has an opening
angle as explained in Figure \ref{kt_eri_disk_config}b and c,
and L1 point is not directly impacted by the nova wind.

In the nova wind phase, the surface layer of the disk is blown
in the wind (e.g., Kelvin-Helmholtz instability) and its photospheric
surface can extend over the Roche lobe size of the WD.
Such examples had been calculated to reproduce the light curves of SSSs
\citep[e.g.,][]{hac03ka, hac03kb, hac03kc}.
A strong support for the existence of a large accretion disk appeared
in the eclipse analysis of the U Sco 2022 outburst.  \citet{mura24ki} 
obtained a large disk (optically thick part)
which is extended up to the L1 point, that is, the size is
$\alpha \sim 1.3$ times the effective Roche lobe
radius\footnote{The effective Roche lobe radius is
defined by the radius of a sphere of which the volume is the same as that
of the inner critical Roche lobe.  We adopt an approximate description 
proposed by \citet{egg83}.} of the WD during the nova wind phase.
The disk (optically thick part) shrinks down to $\alpha \sim 0.85$ times
the effective Roche lobe radius after winds stop.
Here, $\alpha$ is the parameter that represents
the size of the disk in units of the effective Roche lobe radius
(see Equation (\ref{disk_radius_alpha})).
This issue will be described in more detail
in Appendix \ref{large_irradiated_disk_binary}.

Assuming a large disk (Figure \ref{kt_eri_config}a),
the size of which is 1.3 times the effective Roche lobe radius
($\alpha=1.3$), we calculated the model light curve,
as shown by the thick magenta line in Figure 
\ref{kt_eri_only_v_x_big_disk_4500k_logscale}b, which
can be written by the summation of the free-free emission luminosity and
$V$ band fluxes of the photospheric luminosities, i.e.,
\begin{eqnarray}
L_{V, \rm total} &=& L_{V, \rm ff,wind} + L_{V, \rm ph, WD} \cr
  & & + L_{V, \rm ph, disk} + L_{V, \rm ph, comp},
\label{luminosity_summation_wd_disk_comp_v-band}
\end{eqnarray}
where $L_{V, \rm ph, disk}$ is the $V$ band flux from the disk,
and $L_{V, \rm ph, comp}$ the $V$ flux from the companion star. 
The irradiation effect is the main optical source in the disk
and companion star.  See Appendix \ref{irradiated_disk_companion} for
detail of our irradiation calculation.
The thick magenta line in Figure 
\ref{kt_eri_only_v_x_big_disk_4500k_logscale}b
shows a much slower decline than the black line,
and broadly follows the $V,y$ light curve until the SSS phase started.

We have examined the dependency of the model $V$ light curve on the parameter
$\alpha$.  Figure \ref{kt_eri_only_v_x_big_disk_4500k_logscale}b
shows the two cases of  $\alpha=$ 1.3 and 0.9. 
Here, $\alpha=$ 1.3 (magenta line) is taken from \citet{mura24ki}'s
result for U Sco, and $\alpha=$ 0.9 (cyan-blue line) is close to the
tidal limit of an accretion disk.  The two light curves are very similar
because the large orbital period of KT Eri (2.6 days) already allows a 
geometrically large disk even for $\alpha= 0.9$.
The irradiated large disk substantially contributes to the total brightness.
If we further increase $\alpha$, the model $V$ light curve hardly become
brighter than the case of $\alpha= 1.3$.
We interpret that the quick change in the size of the disk is caused by
the quick weakening (or stop) of optically-thick winds.  This issue will be
discussed in Section \ref{transition_wind_to_sss}.


\begin{figure}
\epsscale{1.15}
\plotone{f5.eps}
\caption{
Our light curve model of the disk around the WD of KT~Eri in Figure 
\ref{kt_eri_only_v_x_big_disk_4500k_logscale}b.
The masses of the WD and Roche-lobe-filling companion star
are $1.3 ~M_\sun$ and $1.0 ~M_\sun$, respectively.
The orbital period is $P_{\rm orb}= 2.616$ days.
The inclination angle of the binary is $i=41.2\arcdeg$.
The separation is $10.55 ~R_\sun$ while their effective
Roche lobe radii are $4.24 ~R_\sun$ and  $3.76  ~R_\sun$, respectively.
We assume, in panel (a),
the size of the disk to be 1.3 times the Roche-lobe radius of the WD
and the height of the disk edge to be 0.05 times the disk size
during the nova wind phase but, in panel (b), the disk size is
0.9 times the Roche-lobe radius and the edge height is 0.3 times
the disk size after the winds stop.  The photospheric surfaces of
the disk and companion star are irradiated by the
central hot WD and such irradiation effects are all included in the
calculation of the $V$ light curve reproduction \citep[see][for the
partition of each surface and calculation method of irradiation]{hac01kb}.  
Note that, in panel (a), the disk surface flow matter in Figure
\ref{kt_eri_disk_config} shapes an optically-thick large disk.
A segment of mesh surface on the disk represents the photospheric 
surface of optically-thick part of the large disk. 
Gas is optically thin outside the mesh surfaces (i.e., photospheres
of the disk, WD, and companion star).
In panel (b), a photospheric surface of the disk shrinks inside the Roche
lobe after the nova wind stops.  The L1 stream impacts the disk edge and
makes a spray, which elevates the disk edge \citep{sch97mm}.
We also include the effect of viscous heating in the accretion disk
for a given mass-accretion rate \citep{hac01kb}.
\label{kt_eri_config}}
\end{figure}

\subsection{First plateau in the optical light curve}
\label{Vy_light_curve_mid_late}

Figure \ref{kt_eri_only_v_x_big_disk_4500k_logscale}b shows that 
the $V,y$ brightnesses keep almost a flat plateau in the SSS phase
(day $\sim 80$--240).  Because the wind from the WD weakened and
stopped by day $\sim 100$ in our $1.3 ~M_\sun$ WD model (solid thick black
line), the large disk comes back to a normal size, i.e.,
the edge of the disk shrinks inside the Roche lobe \citep[see, e.g.,
Figure 5 of ][]{mura24ki}.
Assuming a disk size of 0.9 times the effective Roche lobe radius,
we calculated the model
light curve, as shown by the thick magenta line in Figure 
\ref{kt_eri_only_v_x_big_disk_4500k_logscale}b.
This thick magenta line reproduces a flat plateau until the hydrogen
burning ends on day $\sim 240$. 

The end day of hydrogen shell-burning can be detected from 
the quick decay of \ion{He}{2} 4686\AA\  line intensity as well as the
rapid decay of soft X-ray count rate.
We plot the intensities of \ion{He}{2} 4686\AA\  line (magenta line) and
continuum (blue line) in Figure \ref{vflux_o3_he2}b, which are
calculated from the SMARTS spectra of KT Eri \citep{wal12bt}.
Its flux quickly decays from day $\sim 240$ to day $\sim 260$.
This indicates that the high energy photons, which excite \ion{He}{2}
line, quickly decreases from day $\sim 240$ to day $\sim 260$.
Thus, hydrogen-shell burning ended on day $\sim 240$ 
and its soft X-ray or UV flux started to decay.

Such an optical plateau in the SSS phase is frequently observed
in recurrent novae like in U Sco, the origin of which is explained by the 
contribution of an irradiated large accretion disk \citep[e.g.,][]{hkkm00}.  
An optical plateau, however, rarely appears in classical novae. 
Many classical novae are short (a few to several hours) orbital period 
binaries \citep[e.g.,][]{schaefer22a} and, thus, they have
a small size disk that hardly contributes to the optical magnitude.
Among classical novae, KT Eri is a rare exception of
a long (as long as 2.6 days) orbital period \citep{schaefer22wh},
and could have a large disk.  The disk model in the SSS phase
will be explained in more detail
in Appendix \ref{large_irradiated_disk_binary}.

\subsection{Second plateau: high mass-accretion rate}
\label{Vy_light_curve_second_plateau}

At the end of the nova outburst, the optical magnitude quickly decays
with the X-ray flux because hydrogen burning ends and the bright
irradiation effect disappears, although our model includes
irradiation effects of the accretion disk 
and companion star all the time during our simulation.
To explain the brightness of the second
plateau in KT Eri ($V\sim 14.1$), we assume a high mass-accretion rate
of $\dot{M}_{\rm acc}=2\times 10^{-7} ~M_\sun$ yr$^{-1}$, similar to the
value estimated by \citet{schaefer22wh}.  
Note that Schaefer et al.'s estimate of $\dot{M}_{\rm acc}= 3.5\times 10^{-7} 
~(d/5.1{\rm ~kpc})^2 ~M_\sun$ yr$^{-1}$ is converted to
$\dot{M}_{\rm acc}=2\times 10^{-7} ~M_\sun$ yr$^{-1}$ for $d= 4.2$ kpc.
The model light curve (thick
magenta line in Figure \ref{kt_eri_only_v_x_big_disk_4500k_logscale}b)
shows the case of such a high mass-accretion rate.  We include
the effect of viscous heating, the method of which is described
in \citet{hac01kb}.  Viscous heating
in the accretion disk contributes to the brightness of $V\sim 14.1$
(see Table \ref{2nd_plateau_parameters}).

Our assumed mass-accretion rate of $\dot{M}_{\rm acc}=2\times 10^{-7} ~M_\sun$
yr$^{-1}$ is close to the lowest mass-accretion rate for steady hydrogen
burning, $\dot{M}_{\rm steady}\approx 3\times 10^{-7} ~M_\sun$ yr$^{-1}$,
for a $1.3 ~M_\sun$ WD \citep[see, e.g.,][for recent estimates]{wol13bb,
kat14shn}.  If such a high mass-accretion rate continues in the SSS phase, 
hydrogen burns longer because new fuel is supplied.
The SSS phase can be extended up to day $\sim 240$, 
which is consistent with the duration of the observed SSS phase.
If we assume a much smaller mass accretion rate of
$\dot{M}_{\rm acc}=1\times 10^{-9} ~M_\sun$ yr$^{-1}$,
on the other hand, the hydrogen burning stops earlier on day $\sim 180$
(red line in Figure \ref{kt_eri_only_v_x_big_disk_4500k_logscale}b)
and the $V$ brightness is $V\sim 15.3$ at the second plateau
(Table \ref{2nd_plateau_parameters}).
Thus, we can explain both the brightness in the second plateau ($V\sim 14.1$)
and the SSS duration (until day $\sim 240$), at the same time,
with a single mass accretion rate
of $\dot{M}_{\rm acc}=2\times 10^{-7} ~M_\sun$ yr$^{-1}$.

\subsection{Brightness in the quiescent phase: Variable mass-accretion rate}
\label{variable_mass_accretion_rate}

The quiescent brightness of KT Eri is highly variable around at $V\sim 15.1$
\citep[$V=14.1-15.8$:][]{schaefer22wh} after the outburst.
The mean brightness of $V\sim 15.1$ can be reproduced with our model
with the mass accretion rate of 
$\dot{M}_{\rm acc}= 1\times 10^{-8} ~M_\sun$ yr$^{-1}$
(Table \ref{2nd_plateau_parameters}).
We suppose that the mass accretion rate had dropped
from $\sim 2\times 10^{-7} ~M_\sun$ yr$^{-1}$
to $\sim 1\times 10^{-8} ~M_\sun$ yr$^{-1}$ ($V=15.1$) or
to $\sim 1\times 10^{-9} ~M_\sun$ yr$^{-1}$ ($V=15.3$)
at the end of the second plateau phase, as depicted by the blue line
with an arrow in Figure \ref{kt_eri_only_v_x_big_disk_4500k_logscale}b. 

\citet{schaefer22wh} raised the question on the origin of 
a large amplitude variation in quiescence.  Our quick analysis shows that
a brightness variation between $V\sim 14.1$ and $V \sim 15.8$ could be
resulted from a large variation in the mass accretion rate
between $\dot{M}_{\rm acc}= 2\times 10^{-7} ~M_\sun$ yr$^{-1}$
and $\dot{M}_{\rm acc}= 1\times 10^{-11} ~M_\sun$ yr$^{-1}$ (or virtually
zero; see gray line in Figure \ref{kt_eri_only_v_x_big_disk_4500k_logscale}b
and Table \ref{2nd_plateau_parameters}).
We indicate such a variation by the blue line with arrows in 
Figure \ref{kt_eri_only_v_x_big_disk_4500k_logscale}b, although the origin
of this large variation of mass transfer rate is unknown.
We should note
that the photospheric temperature of the companion star to be as low as
4,500 K, which is much lower than \citet{schaefer22wh}'s estimate of 6,200 K.
We discuss more details of the brightness change
in Appendix \ref{brightness_quescent_phase}.

\subsection{Short summary of binary parameters}
\label{short_summary_binary_parameter}

To explain the temporal variation of SMEI, $V$, and $y$ light curves
of KT Eri, we adopt the distance to KT Eri $d=4.2$ kpc,
reddening $E(B-V)=0.08$ \citep{rag09bs},
distance modulus in the $V$ band $(m-M)_V=13.4$,
WD mass $M_{\rm WD}= 1.3 ~M_\sun$,
companion mass $M_2= 1.0 ~M_\sun$ \citep{schaefer22wh},
surface temperature of the companion $T_{\rm ph,2}= 4,500$ K,
separation $a=10.55 ~R_\sun$, Roche lobe effective radii
$R_{\rm RL,1}= 4.24 ~R_\sun$ and $R_{\rm RL,2}= 3.76 ~R_\sun$, 
inclination angle $i= 41.2\arcdeg$ 
for $K_{\rm WD}= K = 58.4$ km s$^{-1}$ \citep{schaefer22wh},
and orbital velocities $v_1= 88.68$ km s$^{-1}$ and
$v_2= 115.3$ km s$^{-1}$, and
orbital period $P_{\rm orb}= 2.616$ days \citep{schaefer22wh}.
The mass transfer rate from the companion star keeps a high value of
$\dot{M}_2\sim -2\times 10^{-7} ~M_\sun$ yr$^{-1}$
until the second plateau ends, and then sometimes decreases down to
virtually zero (no mass accretion), sometimes goes up to
$\dot{M}_{\rm acc}\sim 1\times 10^{-7} ~M_\sun$ yr$^{-1}$, but its average is
$\dot{M}_{\rm acc}\sim 1\times 10^{-8} ~M_\sun$ yr$^{-1}$ for $V \sim 15.1$,
or $\dot{M}_{\rm acc}\sim 1\times 10^{-9} ~M_\sun$ yr$^{-1}$
for $V \sim 15.3$.


The surface of the accretion disk is blown in the wind.
We also adopt the assumption that its optically
thick part of the surface could be extended up to close to the L1 point,
that is, 1.3 times the Roche lobe size during the nova wind phase,
similar to what was found observationally for U Sco.
When nova winds stop, its outer size shrinks
down to 0.9 times the Roche lobe size, close to the tidal limit of
an accretion disk.
Then, the bright irradiation effects of the 
disk and companion star reproduce the brightness of the first
plateau ($V\sim 12.0$) and the viscous heating of the disk 
($\dot{M}_{\rm acc} \sim 2\times 10^{-7} ~M_\sun$ yr$^{-1}$) 
explains the brightness ($V\sim 14.1$) in the second plateau.   

\section{Discussion}
\label{discussion}

\subsection{Is KT Eri a recurrent nova?}
\label{kt_eri_recurrent_nova_no}

\citet{schaefer22wh} estimated the mass-accretion rate to the WD 
to be $3.5\times 10^{-7} ~(d/5.1{\rm ~kpc})^2 ~M_\sun$ yr$^{-1}$
from the brightness of an accretion disk and concluded that
KT Eri is a recurrent nova with the recurrence time of 40--50 yr.
In this subsection, we examine whether or not their conclusion is
supported by our results. 

Our model $V$ light curves, together with variable $-\dot{M}_2$,
suggest that (1) the mass-transfer rate ($-\dot{M}_2$) is as high as
$2\times 10^{-7} ~M_\sun$ yr$^{-1}$ in the first and second plateaus,
but (2) $-\dot{M}_2$ drops to $1\times 10^{-8} ~M_\sun$ yr$^{-1}$
for $V\sim 15.1$ or to $1\times 10^{-9} ~M_\sun$ yr$^{-1}$ for $V\sim 15.3$
(Table \ref{2nd_plateau_parameters}),
close to an average in the quiescent phase after the outburst.
Here, $\dot{M}_2$ is the mass losing rate (negative) of the companion.
If the total mass is conserved, we have $\dot{M}_{\rm acc} + \dot{M}_2 = 0$.
This lower mass-transfer rate is consistent with the suggestion from
the MMRD position of KT Eri (at $M_{\rm WD} \sim 1.3 ~M_\sun$,  
$\dot{M}_{\rm acc}\sim 1\times 10^{-9} ~M_\sun$ yr$^{-1}$, and
the recurrence time of $\sim 3,000$ yr) in Figure 
\ref{vmax_t3_vmax_t2_selvelli2019_schaefer2018_2fig}.
This decreasing tendency of mass-accretion rate in quiescence 
is also consistent with $\dot{M}_{\rm acc} \approx 1.9 \times 10^{-10} 
~(d/3.7{\rm ~kpc})^2 ~M_\sun$ yr$^{-1}$ observed 
by \citet{sun20od} about 8.3 yr after the outburst.

We should point out that, if we adopt the distance of $d=4.2$ kpc instead
of Schaefer et al.'s $d= 5.1$ kpc, we have the mass-accretion rate of 
$\dot{M}_{\rm acc}= 2\times 10^{-7} ~M_\sun$ yr$^{-1}$ from their
$3.5\times 10^{-7} ~(d/5.1{\rm ~kpc})^2 ~M_\sun$ yr$^{-1}$, which is 
consistent with our estimate during the second plateau.  However,
the average mass-accretion rate in quiescence becomes
$\dot{M}_{\rm acc}= 1\times 10^{-8} ~M_\sun$ yr$^{-1}$, 
corresponding to the brightness variation from
$V\sim 14.1$ (second plateau) to $V\sim 15.1$ (quiescence, average).
Schaefer et al.'s very high $\dot{M}_{\rm acc}$ is probably valid only
for the second plateau, but the mean mass-accretion rate in quiescence
could be as low as $\sim 1\times 10^{-9} ~M_\sun$ yr$^{-1}$ in a long
interoutburst period as suggested by the position of KT Eri in
Figure \ref{vmax_t3_vmax_t2_selvelli2019_schaefer2018_2fig}.  

The estimated recurrence period is directly 
connected to the adopted mass-accretion rate. 
If we adopt the $V$ brightness in quiescence before the outburst,
$V_{\rm q, min}=15.3$, $<V_{\rm q}>=14.5\pm0.2$, and $V_{\rm q, max}=14.0$
\citep{schaefer22wh},  
the corresponding mass-accretion rate is calculated to be
$\dot M_{\rm acc}\sim 1\times 10^{-9} ~M_\sun$ yr$^{-1}$, 
$\sim 1\times 10^{-7} ~M_\sun$ yr$^{-1}$, and 
$ \sim 2\times 10^{-7} ~M_\sun$ yr$^{-1}$, as 
in Figure \ref{kt_eri_only_v_x_big_disk_4500k_logscale}. 
These give us the recurrence times of 3600, 10, and 3.7 yr, respectively,
for our $1.3 ~M_\sun$ WD \citep{hac20skhs}.
The estimated recurrence times do not change much
for the other WD masses of $1.28$ and $1.32 ~M_\sun$.
The 10 and 3.7 yr recurrence times are not consistent with 
\citet{jur12rd}'s results.
This indicates that we cannot determine the recurrence time 
only from a short period observation in a much longer interoutburst duration.

\citet{jur12rd} searched the archival plates of the Harvard College
Observatory for a previous outburst of KT Eri, but found no outbursts
between 1888 and 1962, and concluded that, even if KT Eri
is a recurrent nova, it should have a recurrence time of centuries.
This long recurrence time is also consistent with the suggestion 
($\sim 3,000$ yr) from the MMRD position of KT Eri in Figure 
\ref{vmax_t3_vmax_t2_selvelli2019_schaefer2018_2fig} (and the 3600 yr
from $V_{\rm q, min}=15.3$) but does not support
both the 10 yr from $<V_{\rm q}>=14.5\pm0.2$ and the 3.7 yr from
$V_{\rm q, max}=14.0$.
Thus, the mean brightness in a short timescale compared with a long
interoutburst period has nothing on the mean mass-accretion rate to 
determine the recurrence period.
All of these features consistently suggest that KT Eri is not a recurrent
nova but its recurrence time is as long as $\sim 3,000$ yr.

Finally, we compare the timescale of $t_{\rm peak}$, the days from the
outburst to optical maximum, between the observation and a theoretical model. 
We estimated the outburst day of KT Eri, from the rising trend
in the SMEI light curve, to be $t_{\rm peak}= 2.7$ days in Section
\ref{observation_interpretation_kt_eri}.
This estimate is broadly consistent with $t_{\rm peak}= 2.5$ days
for a self-consistent nova model of a $1.3 ~M_\sun$ WD with 
$\dot M_{\rm acc}= 2\times 10^{-9} ~M_\sun$ yr$^{-1}$
\citep[model M13C10; ][]{kat24sh}.  The recurrence time of model M13C10
is about 1000 yr, also suggesting that KT Eri is not a recurrent nova.


\begin{figure*}
\gridline{\fig{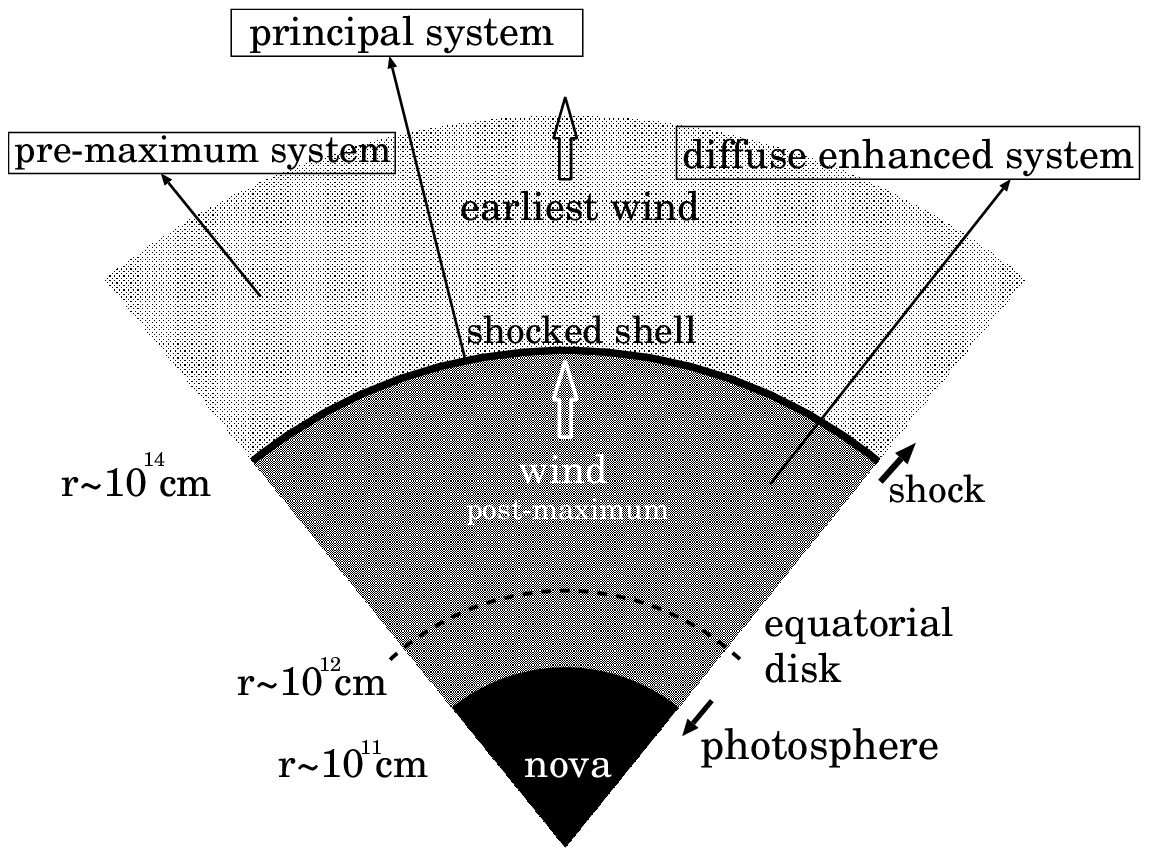}{0.55\textwidth}{(a) day 45-60 (wind phase)}
          \fig{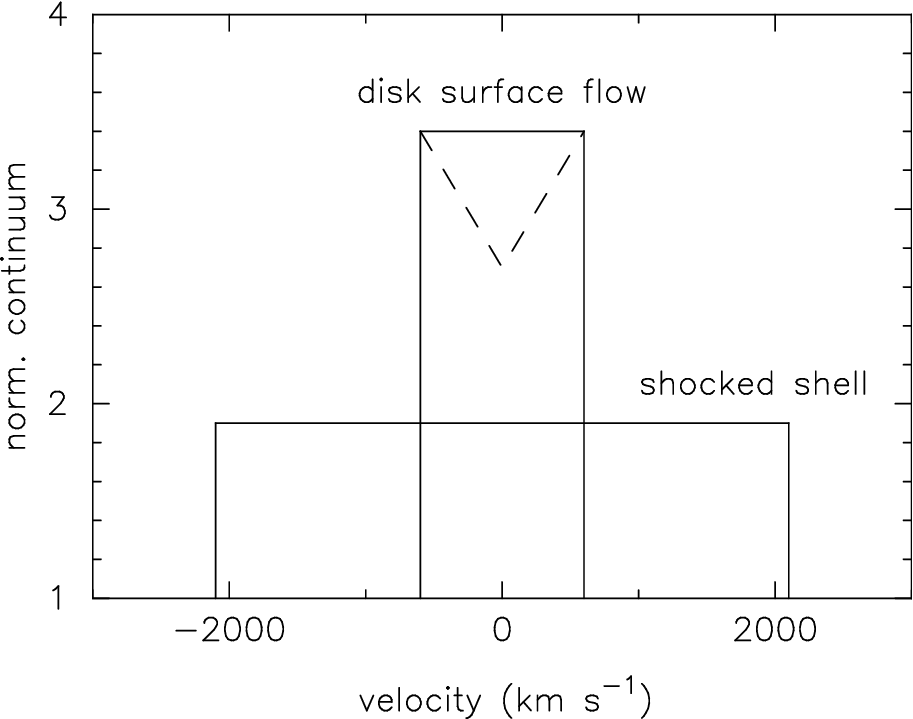}{0.45\textwidth}{(b) H$\alpha$ line profile in the
wind phase}
          }
\gridline{\fig{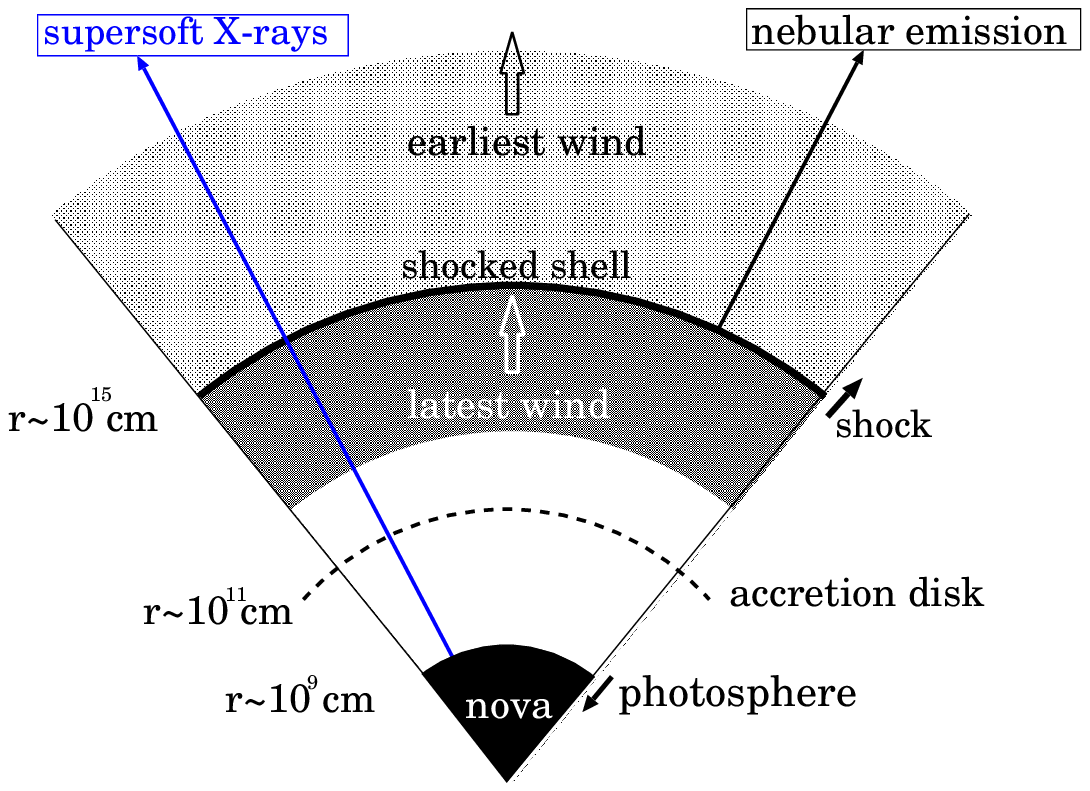}{0.55\textwidth}{(c) day 100-150 (SSS phase)}
          \fig{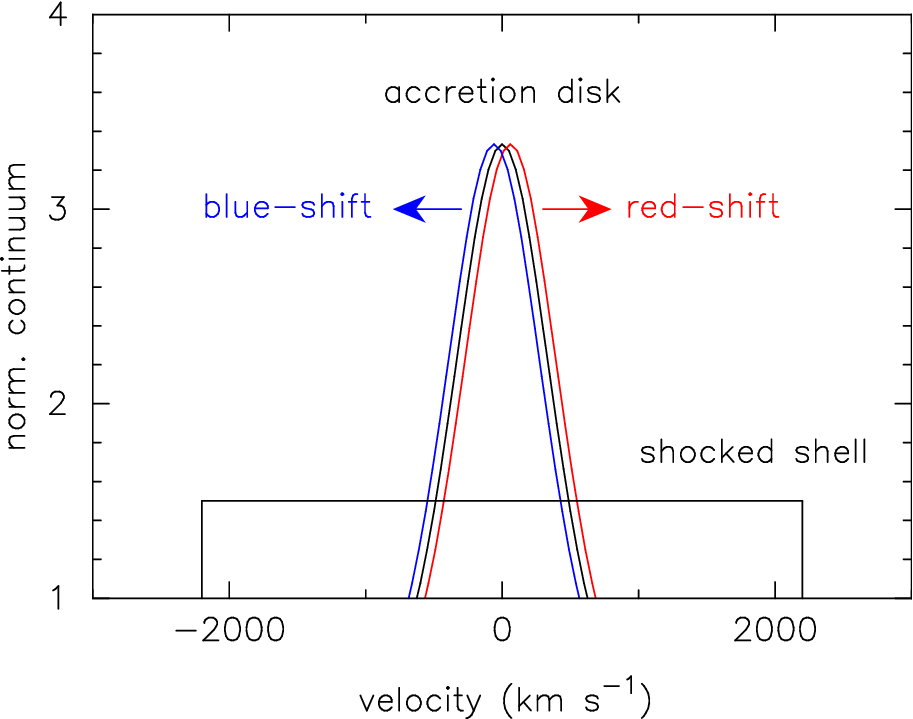}{0.45\textwidth}{(d) H$\alpha$ line profile in the
SSS phase}
          }
\caption{
(a) Schematic illustration of a nova ejecta configuration in the wind phase
of KT Eri (day $\sim 45$ --- 60).  The photosphere of the WD envelope has
already shrunk to $R_{\rm ph}\lesssim 1 ~R_\sun$ and an equatorial disk
appears.  A shock wave arose just after the optical maximum and has already
moved far outside the WD photosphere (and the binary).
The shocked shell emits Balmer lines such as H$\alpha$.
The surface of the disk absorbs ultraviolet photons
and becomes hot to emit H$\alpha$ photons.
This figure is taken from Figure 2(b) of \citet{hac23k} with a modification.
We assume that ejecta are spherically symmetric, but the disk is located
on the orbital plane which is inclined with the angle of $i=41.2\arcdeg$
from the line of sight.
(b) Schematic H$\alpha$ line intensity profiles both from the disk surface
flow (narrower component) and shocked shell (broader pedestal component).
The shocked shell is geometrically thin, optically thin, and spherically
symmetric, so its line profile is approximated by a rectangle with
the line width of $v_{\rm exp}= v_{\rm shock} \sim 2100$ km s$^{-1}$.
The central narrower rectangular component is calculated from the 
outflowing disk surface flow with the velocity of $v_{\rm disk}\sim 900$
km s$^{-1}$ ($= 600$ km s$^{-1} / \sin 41.2\arcdeg$) by assuming
that the emissivity of H$\alpha$ is cylindrically uniform on the disk surface.
These line profiles mimic the H$\alpha$ line profiles observed by
\citet{rib13bd}.  The dashed line mimics the effect of self-absorption.
(c) Schematic illustration of a nova ejecta configuration in the SSS phase
of KT Eri (on day $\sim 100$ --- 150). 
The photosphere of the WD envelope has already shrunk to 
$R_{\rm ph}\lesssim 0.1 ~R_\sun$ and emits supersoft X-rays. 
The shocked shell emits Balmer lines such as H$\alpha$ and other nebular lines.
This figure is taken from Figure 2(d) of \citet{hac23k} with a modification.
(d) Schematic line intensity profiles both from the accretion disk and
shocked shell.  The line profiles from the shocked shell are approximated
by a rectangle with the line width of
$v_{\rm exp}= v_{\rm shock} \sim 2,200$ km s$^{-1}$.
The central narrow line profiles mimic the line profile 
of the accretion disk with FWHM$\sim 1000$ km s$^{-1}$ 
\citep[taken from the H$\alpha$ line width in][]{mun14mv}.
The line center moves along with the orbital motion.
\label{kt_eri_line_profile}}
\end{figure*}


\subsection{Temporal changes of emission line profiles}
\label{emission_line_profile}

We assumed a large disk during the nova wind phase in KT Eri 
(Section \ref{observation_interpretation_kt_eri} and Appendix
\ref{large_irradiated_disk_binary}).
Here, we discuss whether or not the temporal changes of emission line
profiles in KT Eri is consistent with our disk model in Figures
\ref{kt_eri_disk_config} and \ref{kt_eri_config}.

\subsubsection{Emission line profiles in the wind phase}
\label{emission_line_profile_wind}

\citet{mun14mv} reported spectroscopy of KT Eri between UT 2009 December 1
and UT 2014 February 14 (from day 19 to day 1553). 
They discussed the appearance and evolution of a narrow \ion{He}{2} 4686\AA\  
emission line as well as Balmer lines of H$\alpha$ and H$\beta$.
These lines show a narrow emission profile 
superimposed to much broader emission components.
Figure \ref{kt_eri_line_profile}b shows a schematic illustration of
such two narrow and broad components.

Similar line profiles have been observed in other several novae, YY Dor,
Nova LMC 2009, U Sco, DE Cir, and V2672 Oph \citep[e.g.,][for
spectra]{mun11rb, mas12ew, mas14m, tak15d}.  These complex line profiles
have always been modeled with asymmetric ejecta geometries, consisting of
bipolar lobes, polar caps, and equatorial rings \citep[see ][for
KT Eri]{rib13bd}.  

Instead, in the nova wind phase, we interpret that the above narrow
and broad line profiles come from a disk surface flow and spherically
symmetric shocked shell, respectively.

\citet{kat22sha} computed a nova outburst evolution including self-consistent
wind mass-loss.  Based on this result, \citet{hac22k} calculated
a strong shock formation far outside the nova photosphere,
and presented a configuration of the nova ejecta after the shock arose
(Figure \ref{kt_eri_line_profile}a).
The shocked shell divides the ejecta into
three parts, the earliest wind (before optical maximum), shocked shell,
and inner wind.  \citet{hac22k} interpreted that these three parts
contribute to the three emission/absorption line systems defined by 
\citet{mcl42}, respectively,
as illustrated in Figure \ref{kt_eri_line_profile}a.

The shocked shell is a bright source of H$\alpha$, because the shocked
shell eventually collects $\sim 90$\% of the nova ejecta mass
\citep{hac22k}. 
If the shocked shell is spherically symmetric, geometrically thin,
and optically thin, the H$\alpha$ profile is approximated by
a rectangle \citep{bea31} with the width of $v_{\rm shell}=
v_{\rm shock}\approx 2100$ km s$^{-1}$, as illustrated in Figure 
\ref{kt_eri_line_profile}b.  Thus, the broader pedestal component
should originate from the shocked shell with the velocity of
$v_{\rm shell}\sim 2100$ km s$^{-1}$ in KT Eri.

It has been discussed that the sphericity (or elongation) of a nova shell
depends on its (nova) speed class.  The faster the nova speed class is,
the more spherical the shell is \citep[e.g.,][]{dow00}.
KT Eri belongs to the very fast novae class ($t_2=6.6$ days) and
its shell is probably close to spherical. Therefore, our assumption
of spherical symmetry does not so largely depart from the true shell.

The narrow components might arise from a disk surface flow 
in Figure \ref{kt_eri_disk_config}c.
This is because a thin surface layer of the disk is blown in the wind and
its surface boundary flow has a smaller velocity than that of the wind
itself (see more details in Appendix \ref{disk_not_disrupted}).
We calculated the narrower line profile, 
as in Figure \ref{kt_eri_line_profile}b,
a rectangle with the width of
$v_{\rm incline}\approx 600$ km s$^{-1} = 900$ km s$^{-1} \times
\sin 41.2\arcdeg$, assuming a velocity of the disk surface flow
in the radial direction, $v_{\rm disk}= 900$ km s$^{-1}$,
where we adopt the inclination angle of $i=41.2\arcdeg$
for the binary consisting of a $1.3 ~M_\sun$ WD and $1.0 ~M_\sun$ companion
with the orbital period of $P_{\rm orb}= 2.616$ days.
We further assume circularly uniform emissivity distribution
of H$\alpha$ on the disk surface. 
The concave or dip feature on the top could be due to self-absorption
(dashed line in Figure \ref{kt_eri_line_profile}b)
because the boundary layer of disk surface flow is optically thick 
in the sense of line opacity, where H$\alpha$ lines come from. 

Our narrow and wide component model for H$\alpha$ emission lines
approximately reproduces the H$\alpha$ line profiles in KT Eri 
\citep[see, e.g., Figures 2, 4--7 of ][]{rib13bd}.

\subsubsection{Emission line profiles in the SSS phase}
\label{emission_line_profile_sss}

Figure \ref{kt_eri_line_profile}c depicts a schematic configuration of
the nova ejecta in the SSS phase of KT Eri.
The optically-thick wind had already stopped emerging from the photosphere,
but its tail still collides with the shock,
and the WD photosphere emits supersoft X-ray photons.  

Figure \ref{kt_eri_line_profile}d shows a schematic illustration of
the line profile in such a SSS phase: the line consists of a narrower
Gaussian-like component and a broader pedestal component.
The broader pedestal component originates from
the shocked shell, as in the wind phase.
A strong shock arises outside the nova photosphere \citep{hac22k, hac23k},
and propagate far beyond the binary orbit.   Thus, the broader pedestal
component of H$\alpha$ emission line hardly depends on the orbital phase. 
If the shocked shell is spherically symmetric, geometrically thin,
and optically thin, the H$\alpha$ profile is a rectangle \citep{bea31} 
with the width of $v_{\rm shell}= v_{\rm shock}= 2,200$ km s$^{-1}$,
as illustrated in Figure \ref{kt_eri_line_profile}d.
Here, we assume that the shocked shell was accelerated from
$v_{\rm shell}= 2,100$ km s$^{-1}$ in the wind phase
to $v_{\rm shell}= 2,200$ km s$^{-1}$ in the SSS phase.
See also the doublet [\ion{O}{3}] line profiles of KT Eri obtained on day 294 
by \citet{ara13ii}, each component of which shows a broad 2,200 km s$^{-1}$
rectangular (pedestal) shape in the second plot of their Figure 2.

The narrow components might arise from the accretion disk of the binary
\citep[e.g.,][]{sek88fw, sek89cf, tho01dl, wal11b} while a different
interpretation on the narrow component of U Sco was proposed by \citet{tak15d}.
Here, we assume that the narrow component originates from the accretion disk
because the disk is very bright due to an irradiation effect by the
central hot WD (see more details
in Appendix \ref{large_irradiated_disk_binary}).
Although no broader pedestal components were observed,
similar properties of narrow components of \ion{He}{2} and Balmer emission
lines are also observed in several persistent SSSs 
\citep[e.g.,][]{cow98sc,  sch00ct}.
Such a Gaussian-like line could be blue- or red-shifted by the orbital
motion, as illustrated in Figure \ref{kt_eri_line_profile}d, because
narrow Gaussian-like \ion{He}{2} or H$\alpha$ lines come from
the irradiated accretion disk.

\subsubsection{Transition from the wind phase to SSS phase}
\label{transition_wind_to_sss}

\citet{mun14mv} pointed out that the drastic change in the narrower component
occurred on day $\sim 70$ from like Figure \ref{kt_eri_line_profile}b to
like Figure \ref{kt_eri_line_profile}d.  This can be explained
if the disk configuration changes from Figure \ref{kt_eri_config}a to
Figure \ref{kt_eri_config}b around on day $\sim 70$ (when the wind
drastically weakened).  
In fact, Figure \ref{vflux_o3_he2}b shows the temporal variation of 
\ion{He}{2} 4686\AA\  line intensity, in which
the intensity (magenta line) quickly drops from $\sim 2\times 10^{-11}$ 
to $\sim 2\times 10^{-12}$ erg s$^{-1}$ cm$^{-2}$ \AA$^{-1}$ between
day $\sim 70$ and $\sim 90$.  This 10 times drop in the intensity
corresponds to both the shrinkage of the disk surface area and
the rapid decrease in the wind mass-loss rate.
Thus, our theoretical explanations of \ion{He}{2} and Balmer emission lines
are broadly consistent with \citet{mun14mv}'s observational results.
Such a rapid shrinkage of the disk radius from $\alpha \sim 1.3$ to
$\alpha\sim 0.85$ was observed $\sim 25$ days after optical maximum
in the 2022 outburst of U Sco \citep[see Figure 5 of ][]{mura24ki}.

To summarize, the change from Figure \ref{kt_eri_line_profile}b to
Figure \ref{kt_eri_line_profile}d is composed of two things:
(1) One is that the broader pedestal component shows a gradual increase
in the velocity (acceleration from 2,100 to 2,200 km s$^{-1}$)
but a gradual decrease in the intensity over the transition
from the wind phase to the SSS phase.
This is consistent with our interpretation of shocked shell-origin,
because the shocked shell is located far outside the binary and
not directly related with the motion of binary.
(2) The other is that the line profile of the narrower component 
drastically changes from the wind phase to the SSS phase.
Moreover, the intensity of \ion{He}{2} 4686\AA\  emission line
drops by a factor of 10 (Figure \ref{vflux_o3_he2}b).
This drop in the intensity is consistent with both the rapid shrinkage
in the surface area of the disk from the large disk to a normal size of
the disk as well as the quick decrease in the wind mass-loss rate
just before winds stop.
Thus, the features of narrow Gaussian-like components of KT Eri
are common among the several persistent SSSs (non novae),
whereas the broader pedestal components appeared only in novae such as KT Eri. 
Persistent SSSs have no strong shocked shell,
but their binary nature is similar to KT Eri in the SSS phase, i.e.,
like in Figure \ref{kt_eri_config}b.

\section{Conclusions}
\label{conclusions}
Our main results are summarized as follows:
\begin{enumerate}
\item Our $1.3 ~M_\sun$ WD (Ne3) model well reproduces the SMEI, $V$,
and $y$ light curves of KT Eri.  In our model, the $V$ light curve
consists of free-free (FF) emission from the wind just outside the
photosphere plus blackbody (BB) emission from the WD photosphere.
The supersoft X-ray flux (0.3-10.0 keV) is calculated from blackbody emission
from the WD photosphere.  The relatively long duration of supersoft
X-ray source (SSS) phase can be explained by a relatively large
mass-transfer rate of $-\dot{M}_2\sim 2\times 10^{-7} ~M_\sun$ yr$^{-1}$
in the SSS phase. 
\item We obtain the distance modulus in the $V$ band
to be $\mu_V\equiv (m-M)_V=13.4\pm0.2$ from the direct model light curve
fitting for our $1.3 ~M_\sun$ WD (Ne3) model.
This gives a distance of $d=4.2\pm 0.4$ kpc toward KT Eri for the observed
redding of $E(B-V)= 0.08$.
\item Both the $V$ and $y$ brightnesses decline almost in the same way even
in the nebular phase. 
To explain the similar behaviors of $V$ and $y$ light curves in the later
phase, we assume a large disk around the WD, temporally extended by strong
nova winds, the radius of which is 1.3 times the effective
Roche-lobe size, similar to what was found observationally for U Sco.
The brightness itself is not so sensitive to the size of the disk if the
size is larger than 0.9 times the Roche lobe radius, because the disk is
already large for such a long orbital period of 2.6 days.
After the winds stop, we assume a normal size of the disk with an elevated
edge.  Both the disk and companion star are irradiated by the hot WD.
The first plateau ($V\sim 12.0$) is reproduced with
a bright disk irradiated by the hydrogen-burning hot WD.
\item The second plateau ($V\sim 14.1$) is explained by a viscous heating
disk together with a high mass-transfer rate of $-\dot{M}_2
\sim 2\times 10^{-7} ~M_\sun$ yr$^{-1}$.  The $V$ brightness drops
to $V\sim 15.1$ (or $V\sim 15.3$) when the mass-transfer rate decreases to 
$-\dot{M}_2 \sim 1\times 10^{-8} ~M_\sun$ yr$^{-1}$ 
(or $-\dot{M}_2 \sim 1\times 10^{-9} ~M_\sun$ yr$^{-1}$) in quiescence. 
\item The mass-accretion rate of $\dot{M}_{\rm acc}=2\times 10^{-7} ~M_\sun$
yr$^{-1}$ is close to the lowest mass-accretion rate for steady hydrogen
burning, $\dot{M}_{\rm steady}\approx 3\times 10^{-7} ~M_\sun$ yr$^{-1}$,
for a $1.3 ~M_\sun$ WD.  If such a high mass-accretion rate continues
in the SSS phase, hydrogen burns longer because new fuel is supplied.
The SSS phase can be extended up to day $\sim 240$, being
consistent with the duration of the observed SSS phase.
Therefore, we can explain both the brightness in the second plateau ($V\sim 
14.1$) and the SSS duration (until day $\sim 240$),
at the same time, with a single mass accretion rate
of $\dot{M}_{\rm acc}=2\times 10^{-7} ~M_\sun$ yr$^{-1}$.
\item A large temporal brightness variation from $V\sim 14.0$ to $V\sim 15.8$
in quiescence can be explained as a variable mass-transfer rate from
$-\dot{M}_2 \sim 2\times 10^{-7} ~M_\sun$ yr$^{-1}$ to
$-\dot{M}_2 \sim 1\times 10^{-11} ~M_\sun$ yr$^{-1}$ 
(virtually zero mass-transfer rate) for the photospheric
temperature $T_{\rm ph,2}\approx 4,500$ K of the companion star with
$M_2\approx 1.0 ~M_\sun$.
\item The recurrence time is difficult to estimate from the $V$ brightness
in quiescence, because its observational period is too short compared with
a much longer interoutburst period. 
Instead, 
we estimate the recurrence period from the position of KT Eri in the MMRD
diagram to be $\sim 3000$ yr.
\item We analyzed the temporal variation of H$\alpha$ line profiles 
based on our nova model as well as our binary model.  The H$\alpha$ line
consists of a broader pedestal component and a narrower component.
The broader component originates from a spherically-symmetric,
shocked shell with the line width of $v_{\rm shell}=2,100$---2,200 km s$^{-1}$.
We interpret the narrower component as arising from
the outflowing disk surface flow with
the velocity of $v_{\rm disk}=900$ km s$^{-1}$ in the wind phase.
In the SSS phase after winds stop, the disk shrinks to a normal size
and the narrower component becomes similar
to that of persistent supersoft X-ray sources,
the line center of which moves along with the orbital motion.
\item We apply the time-stretching method to the $B$, $V$, and $I_{\rm C}$
light curves of KT Eri, and obtain the distance moduli in the $B$, $V$,
and $I_{\rm C}$ bands in Appendix \ref{time-stretching_method}.
The three different distance-reddening relations
of $(m-M)_B= 13.47\pm 0.2$, $(m-M)_V= 13.4\pm 0.2$, and $(m-M)_I= 13.27\pm 0.2$
cross roughly at the distance of $d=4.2$ kpc and $E(B-V)=0.08$, being
consistent with the Gaia eDR3 distance of $d= 4.1^{+0.5}_{-0.4}$ kpc
\citep{bai21rf}, \citet{schaefer22b}' estimate $d=4211^{+466}_{-296}$ pc,
and the observed reddening of $E(B-V)= 0.08$ \citep{rag09bs}.
\item The maximum $V$ brightness of KT Eri is $M_V=-8.0$ for 
$(m-M)_V= 13.4$ and the decline rates in the SMEI ($\approx V$)
light curve are $t_2=6.6$ and $t_3=13.6$ days \citep{hou10bh}.
The positions of KT Eri in the maximum magnitude versus rate of decline
($t_2$-$M_{V,\rm max}$ and $t_3$-$M_{V,\rm max}$) diagrams for classical
novae \citep{hac20skhs} 
are consistent with our $1.3 ~M_\sun$ WD (Ne3) model, of which 
the recurrence period and average mass accretion rate are 
$\sim$3,000 yr and $\sim 1\times 10^{-9} ~M_\sun$ yr$^{-1}$, respectively.
Thus, we may conclude that KT Eri is not a recurrent nova.
\end{enumerate}

\begin{acknowledgments}

We thank
the Variable Star Observing League of Japan (VSOLJ) and
the American Association of Variable Star Observers (AAVSO)
for the optical photometric data of KT Eri and other novae.
We are also grateful to the anonymous referee
for useful comments that improved the manuscript.

\end{acknowledgments}






\appendix

\section{White Dwarf Models of Optical and Supersoft X-ray Light Curves}
\label{opticall_thick_wind_model}

We present multiwavelength light curves based on the optically thick wind
model of novae and constrain the range of possible white dwarf (WD) masses.
\citet{hac06kb}
calculated many free-free emission light curves for novae with various WD
masses and chemical compositions based on \citet{kat94h}'s nova wind model. 
Our model $V,y$ light curve is calculated from the summation of
free-free (FF) emission from the model nova wind and blackbody (BB)
emission from the model WD photosphere \citep{hac15k}.
We name such a light curve model ``FF+BB'' (see Equations
(\ref{free-free_flux_v-band}) and 
(\ref{luminosity_summation_flux_v-band})).
The absolute magnitude of each FF+BB model light curve
has been calibrated with several
novae with a known distance modulus in the $V$ band 
\citep{hac15k,hac20skhs}.  This can be done by fixing
the coefficient $A_{\rm ff}$ in Equation (\ref{free-free_flux_v-band}).
These model light curves have reproduced the decay
trends of various nova light curves. 
We chose appropriate models among the light curve database  
and select ones that fit reasonably to the observational light curve
of KT Eri. 

\begin{figure*}
\gridline{\fig{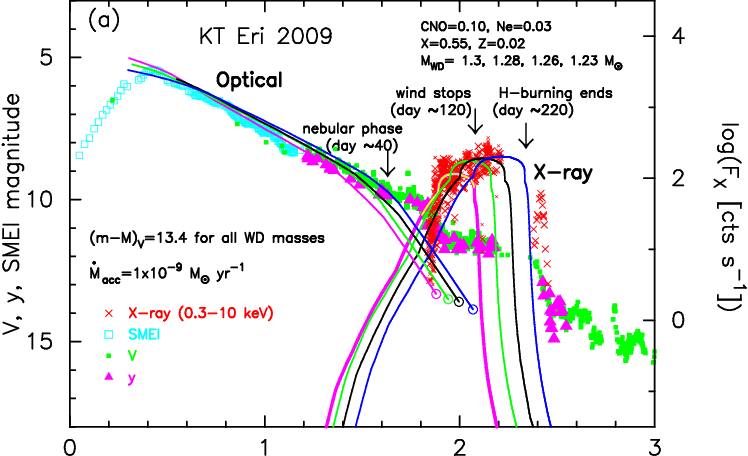}{0.65\textwidth}{}
          }
\gridline{\fig{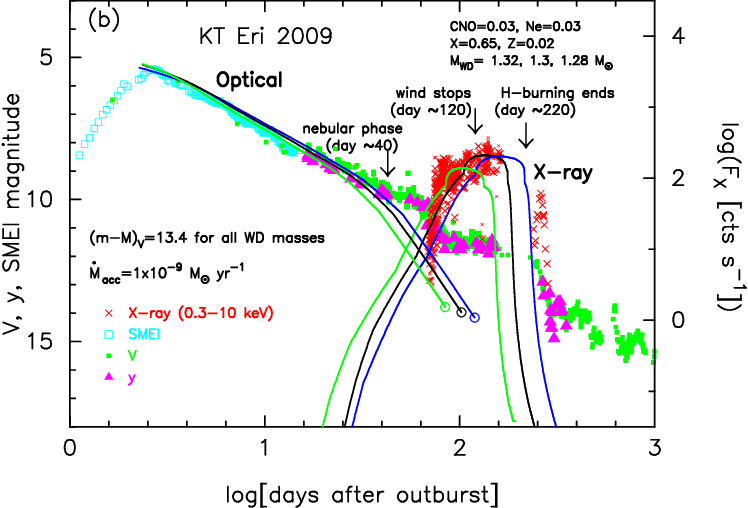}{0.65\textwidth}{}
          }
\caption{
Same as Figure \ref{kt_eri_only_v_x_big_disk_4500k_logscale}b, but
we plot only FF+BB light curves and X-ray (0.3-10.0 keV) light curves
for different sets of WD mass and chemical composition.
The mass-accretion rate onto the WDs
is fixed to be $\dot{M}_{\rm acc}= 1\times 10^{-9} M_\sun$ yr$^{-1}$.
(a) Ne2: $1.23~M_\sun$ (blue line), $1.26~M_\sun$ (black), 
$1.28~M_\sun$ (green), and $1.3~M_\sun$ (magenta but yellow on the top
part of X-ray). 
(b) Ne3: $1.28~M_\sun$ (blue), $1.3~M_\sun$ (black), and $1.32~M_\sun$
(green) WDs.
\label{all_mass_kt_eri_x55z02o10ne03}}
\end{figure*}

Figure \ref{all_mass_kt_eri_x55z02o10ne03} shows our selected models. 
Figure \ref{all_mass_kt_eri_x55z02o10ne03}a
shows the FF+BB model light curves of $1.23~M_\sun$ (blue line),
$1.26~M_\sun$ (black), $1.28~M_\sun$ (green),
and $1.3~M_\sun$ (magenta) WDs
for the chemical composition of Ne2 \citep{hac10k},
i.e., $X=0.55$, $Y=0.30$, $Z=0.02$,
$X_{\rm CNO}=0.10$, $X_{\rm Ne}=0.03$ by mass weight.
Here, $X$, $Y$, and $Z$ are the hydrogen, helium, and heavy element
content, and $X_{\rm CNO}$ is the total abundance of extra carbon,
nitrogen, and oxygen,
and $X_{\rm Ne}$ the extra neon \citep[e.g., ][]{hac06kb}.
These optical fluxes are taken from Table 3 of \citet{hac10k}.
In Figure \ref{all_mass_kt_eri_x55z02o10ne03}b, 
we choose another set of the chemical composition of Ne3, i.e.,
$X=0.65$, $Y=0.27$, $Z=0.02$, $X_{\rm CNO}=0.03$, $X_{\rm Ne}=0.03$
\citep{hac16k}, and plot
three FF+BB model light curves of $1.28~M_\sun$ (blue),
$1.3~M_\sun$ (black), and $1.32~M_\sun$ (green) WDs.
We also add the corresponding X-ray light curves to the figure.
Here, we assume that the mass-accretion rate to the WD is
$\dot{M}_{\rm acc}=1\times 10^{-9} ~M_\sun$ yr$^{-1}$ for all models.

Our model X-ray flux (0.3-10.0 keV) is calculated from the model WD
photosphere with blackbody assumption for the photospheric temperature
$T_{\rm ph}$ and photospheric radius $R_{\rm ph}$ \citep{kat94h}.
Here, we neglect absorption for X-rays outside the photosphere.  It should
be noted that the X-ray flux is not the same as the X-ray count rate,
although the X-ray model light curves broadly catch up the rising phase
of the Swift X-ray count rate.

With the distance modulus in the $V$ band $(m-M)_V=13.4$,
all the model $V$ light
curves broadly follow the $V$ observation of KT Eri until each model 
light curve starts to bend downward.  On the other hand, the rise of
model X-ray light curve broadly catches up with the Swift X-ray observation,
at least, for $M_{\rm WD}=1.3$, $1.28$, and $1.26 ~M_\sun$ (Ne2), in Figure 
\ref{all_mass_kt_eri_x55z02o10ne03}a, but for $M_{\rm WD}=1.32$, $1.3$,
and $1.28 ~M_\sun$ (Ne3), in Figure \ref{all_mass_kt_eri_x55z02o10ne03}b.
Here, we select each WD mass with the $\Delta M_{\rm WD}=0.02 ~M_\sun$ 
interval.  The other WD mass models are excluded (by 
$\Delta M_{\rm WD}=0.02 ~M_\sun$ interval).

Figure \ref{all_mass_kt_eri_x55z02o10ne03} shows, however, that the model
X-ray fluxes decay much earlier than the observed count rate for all of 
these models.  To mitigate the difference, we examined how we extend
the duration of hydrogen burning up to day $\sim 240$, as in Figures
\ref{kt_eri_only_v_x_big_disk_4500k_logscale}b, with high mass-accretion
rates of $\dot{M}_{\rm acc} \gg 1\times 10^{-9} ~M_\sun$ yr$^{-1}$.
This will be described in Section \ref{duration_hydrogen_burning}.


\section{Light curve models of accretion disk and companion star}
\label{large_irradiated_disk_binary}

We describe details of our irradiated disk and binary model for KT Eri.
The large gap between our FF+BB model light curves and the $V,y$ observation
indicates a significant contribution from an irradiated disk.
Here, we assume an accretion disk before nova explosion as in
Figure \ref{kt_eri_disk_config}a, of which 
the surface layer is blown in the nova wind
after nova explosion (Figure \ref{kt_eri_disk_config}b and c).
This is a kind of disk winds in cataclysmic variables,
which is driven by nova winds.   When the disk surface flow (disk wind)
is optically thick, its disk shape looks extending over the Roche lobe
(Figure \ref{kt_eri_config}a).
The photosphere of the large disk emits continuum blackbody flux.
In the optically thick wind phase of a nova, matter is accelerated deep
inside the photosphere.  However, the wind itself becomes optically thin
outside the photosphere.  The density of the wind is rather low compared
with that of the disk.  We plot our model for a nova and accretion disk
in Figure \ref{kt_eri_disk_config}, the details of which are described
in Appendix \ref{disk_not_disrupted}.  The binary parameters and
disk parameters are already described in Section 
\ref{short_summary_binary_parameter}, but we reexamine them in Appendix
\ref{binary_parameter}-\ref{1.3Msun_white_dwarf}.

A bright irradiated disk has been studied in 
the several supersoft X-ray sources and recurrent novae
\citep[e.g.,][]{sch97mm, hkkm00, hac01kb, hac03ka, hac03kb, hac03kc}. 
When the disk is irradiated by the central hot WD, its surface
temperature increases, and each part of the disk surface 
emits photons at this increased temperature.   In our modeling,
the size and shape of the disk are parameters.  We also include
the effect of viscous heating in the accretion disk \citep{hac01kb}.
Our model of such an irradiated disk and companion star is given in
Appendix \ref{irradiated_disk_companion} and the composite model 
light curves are described in Appendix
\ref{temporal_variation_light_curve}-\ref{1.3Msun_white_dwarf}.

\subsection{Disk is not disrupted by nova winds}
\label{disk_not_disrupted} 


\citet{kat22sha, kat24sh} presented nova outburst models that
self-consistently include optically-thick winds as a mass-loss mechanism.
We here show the reason why an accretion disk is not disrupted
in a nova outburst based on \citet{kat24sh}'s nova model.  In Figure
\ref{kt_eri_disk_config}, we show a schematic illustration of 
Kato et al.'s evolution model of a nova envelope for a $1.3 ~M_\sun$ WD,
which is taken from their Figure 6.
After hydrogen ignites at the bottom of the envelope
(Figure \ref{kt_eri_disk_config}a),
the nova envelope expands almost hydrostatically until winds begins to
emerge from the photosphere (Figure \ref{kt_eri_disk_config}b).
This phase can be detected as an X-ray flash because the temperature
of the photosphere is high enough to emit supersoft X-rays.
Such an X-ray flash was first detected by the SRG/eROSITA in the 2020
nova outburst of YZ Ret \citep{kon22wa}.  The X-ray flash of YZ Ret is
consistent with the nova outburst model of a $\sim 1.3 ~M_\sun$ WD 
\citep{kat22shb, kat22shc}.

From the X-ray spectrum analysis of YZ Ret, \citet{kon22wa} found no major
intrinsic absorption during the X-ray flash.
Therefore, \citet{kat22shb} concluded that \\
(1) no dense matter exists around the WD photosphere,\\
(2) no indication of a shock wave, and\\
(3) the hydrogen-rich envelope is almost hydrostatic.\\
Note that all of these are consistent with \citet{kat24sh}'s
theoretical model.  We adopt their conclusions:
optically-thick winds do not yet emerge from the photosphere
at the epoch of X-ray bright (X-ray flash) phase
and the envelope is almost hydrostatic.
Thus, we conclude that the accretion disk is almost intact until
winds begin to emerge from the photosphere (Figure \ref{kt_eri_disk_config}b)
because the envelope expands almost hydrostatically and the density of
the accretion disk is much higher than that of the envelope.

With such information, we assume that the disk in the KT Eri system
is also surviving during the nova outburst and a thin surface layer
of the disk is blown in the wind.  We call such a flow of the boundary
layer between winds and disk the ``disk surface flow.''
The dark-cyan arrows indicate the disk surface flow, the velocity of
which should be slower than that of a wind
(Figure \ref{kt_eri_disk_config}b and c).
The outflowing disk matter shapes a large optically-thick disk.

The nova envelope further expands and, at the same time, blows winds
(Figure \ref{kt_eri_disk_config}c).  The wind is accelerated deep inside
the photosphere.  For this reason, it is called the ``optically thick wind,''
but winds become optically thin outside the photosphere as depicted
in Figure \ref{kt_eri_disk_config}c.
The velocity of wind at the photosphere decreases as the photosphere expands
\citep[see Figure 6 of ][]{kat24sh}.
The wind mass-loss rate also reaches maximum at maximum expansion of
the photosphere.  At the same time,
the free-free emission brightness also attains its maximum (optical maximum)
because of Equation (\ref{free-free_flux_v-band}).
A substantial part of the envelope is blown in the wind.
We plot the envelope structure at/near maximum expansion of photosphere
in Figure \ref{kt_eri_disk_config}c.

The winds are accelerated by the radiation pressure gradient
deep inside the photosphere. We call this region the ``acceleration region
of winds.''  The thick light-cyan circular lines in 
Figure \ref{kt_eri_disk_config}b and c correspond to this acceleration
region of winds.  The envelope is almost static (very low velocity)
inside the acceleration region \citep{kat22sha, kat24sh}.
Therefore, we may conclude that 
the accretion disk is almost intact inside the acceleration region.
The radius of the acceleration region is about $r_{\rm accel}\sim 0.1 ~R_\sun
\sim 10^{10}$ cm when winds start to emerge from the photosphere but expands
up to $r_{\rm accel}\sim 1 ~R_\sun \sim 10^{11}$ cm at maximum expansion
of the photosphere (Figure \ref{kt_eri_disk_config}c).
The momentum of wind is spherically radial so that it does not so push
the accretion disk outward because it has already an open angle avoiding
the disk inside the acceleration region.
Thus, we suppose that the accretion disk is almost intact, at least,
from the view point of spherically radial momentum. 

\citet{dra10o} reported their 3D numerical calculations on the 2010 nova
outburst of U Sco and concluded that the accretion disk
is completely destroyed by the blast wave.  However, \citet{mura24ki}
observationally showed that a large size of the optically-thick
light source exists during the nova wind phase,
about 14 days after optical maximum, in the 2022 outburst of U Sco.
They showed that the light source edge was extended close to the L1 point.
They interpreted this as a structure change of the disk, the size of which
is expanded to $\sim 1.2$ times 
the effective Roche lobe radius of the white dwarf component.
The size of the optical source drastically shrinks to $\sim 0.8$
times the effective Roche lobe radius 26 days after optical maximum,
close to the tidal limit of an accretion disk. 
This observational result and its interpretation
are consistent with our model that only the 
outermost surface region of the disk is blown in the wind (disk surface flow)
and the outer edge of the optically thick part of the disk is extended
over the Roche lobe size \citep[see, e.g., ][]{hac03ka, hac03kb, hac03kc}. 

\citet{dra10o} set up their initial model by a spherical Sedov-type shock
wave at the radius of $r_0= 4\times 10^5$ km $= 4\times 10^{10}$ cm with
the total energy of $E_0\sim 10^{44}$ erg.  This initial set-up is not
supported by \citet{kat24sh}'s nova outburst model as shown in 
Figure \ref{kt_eri_disk_config}c.  \citet{hac22k} showed that
a strong shock arises only after optical maximum (after maximum
expansion of the photosphere).   When the shock arises, 
this radius already expands to, typically,
$r_0\sim 10^{12}-10^{13}$ cm or more.
The separation of U Sco binary is about $a= 6.5 ~R_\sun\sim 
0.5\times 10^{12}$ cm, which is comparable to or smaller than
the formation radius of a shock.
In other words, a shock does not collide with the disk
because the shock formation radius is outside the disk edge in the
U Sco binary system.
\citet{hac22k} also showed that, even if a strong shock arises
and collides with the disk, it is a reverse shock, different from
the Sedov-type shock.
The reverse shock appears when the inner fast moving matter collides
with the outer slow moving matter, which therefore depends
on the faster wind inside the shock 
\citep[see, e.g., Figure 4 of ][]{hac22k}.
The inner wind has an open angle avoiding the disk, so that
no spherical reverse shock arises in the accretion disk. 
  
Thus, the numerical results by \citet{dra10o} on the recurrent nova U Sco
are not supported by both the observation \citep{mura24ki}
and the nova outburst models \citep{hac22k, kat22sha, kat24sh}.
Therefore, the dynamical impact of the fast outflow
(optically thick wind) does not disrupt the disk in the early phase
(until optical maximum).

We do not include any other effects that
would be important for disk disruption. For example, a prolonged contact
with the hot, quasi-hydrostatic envelope could lead to a thermal evaporation
of a significant amount of disk material.

We have already assumed that the
disk surface flows (disk winds) strip off a very thin surface layer of
the disk.  We suppose that this effect is the largest mass loss from
the accretion disk (Figure \ref{kt_eri_disk_config}b and c).
Unfortunately, we cannot correctly estimate the mass-loss rate by disk surface
flows.  At the present time, therefore, we simply assume that the optically
thick part of the disk surface flows remains during the nova wind phase
and this large disk (shape) contributes to the $V$ and
$y$ magnitudes as shown in Figure 
\ref{kt_eri_only_v_x_big_disk_4500k_logscale}b.
If we successfully reproduce the $V$ and $y$ magnitudes by this assumption
of the large disk, we may conclude that a large disk could survive
during the nova outburst of KT Eri.

\subsection{Binary parameters}
\label{binary_parameter} 

We have already obtained a parameter set of 
$M_{\rm WD}= 1.3 ~M_\sun$ (Ne3) and $T_{\rm ph,2}= 4,500$ K for
$M_2= 1.0 ~M_\sun$ in Section \ref{short_summary_binary_parameter},
which reasonably well reproduced the $V,y$ optical light curves
as well as the X-ray light curve.  However, we start here 
$M_{\rm WD}= 1.28 ~M_\sun$ (Ne3) and $T_{\rm ph,2}= 6,200$ K for
$M_2= 1.0 ~M_\sun$, based on
\citet{schaefer22wh}'s suggestion ($M_{\rm WD}= 1.25\pm 0.3 ~M_\sun$
and $T_{\rm ph,2}= 6,200$ K for $M_2= 1.0\pm 0.2 ~M_\sun$).  
For this set of parameters, the separation of the binary is 
$a= 10.5 ~R_\sun$, the effective
radii of each Roche lobe are $R_{\rm RL,1}= 4.2 ~R_\sun$ and
$R_{\rm RL,2}= 3.8 ~R_\sun$, the orbital velocity of the WD is 
$v_{\rm orb,1}= 89.2$ km s$^{-1}$, for the orbital period of 
$P_{\rm orb}= 2.616$ days \citep{schaefer22wh}.
We estimate the inclination angle of the 
binary to be $i= 40.9\arcdeg$ assuming that $K=58.4$ km s$^{-1}$ is coming
from the orbital velocity of the WD \citep{schaefer22wh}.   

In what follows, we change $M_{\rm WD}$ and $T_{\rm ph,2}$ and
examine whether or not the combination of $M_{\rm WD}$ and
$T_{\rm ph,2}$ reproduces the $V$,$y$ and X-ray light curves of KT Eri,
as shown in Figure \ref{kt_eri_only_v_x_big_disk_4500k_logscale}b.


\begin{figure*}
\gridline{\fig{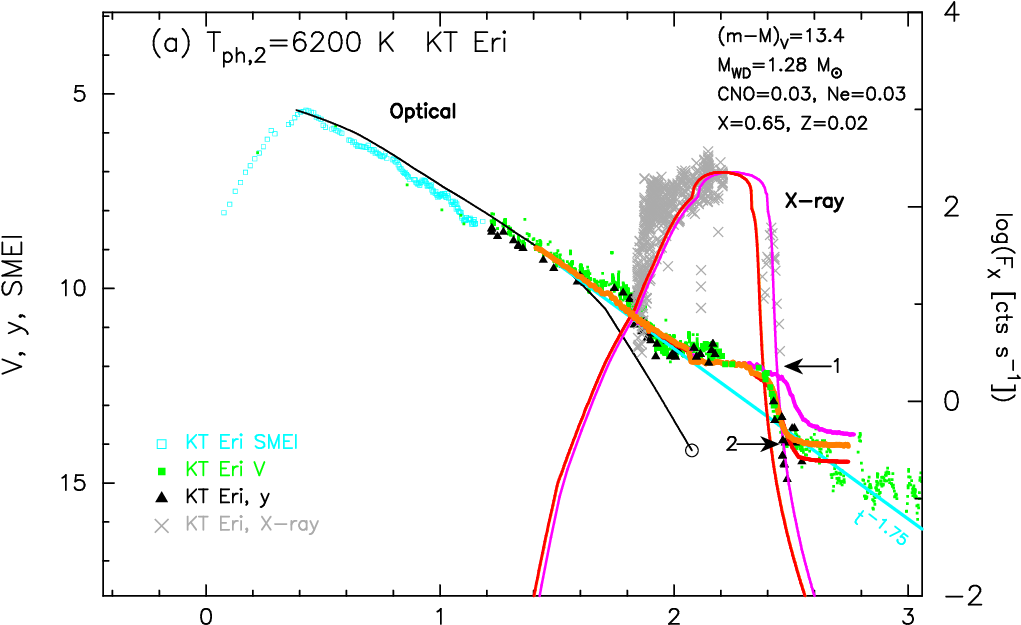}{0.75\textwidth}{}
          }
\gridline{\fig{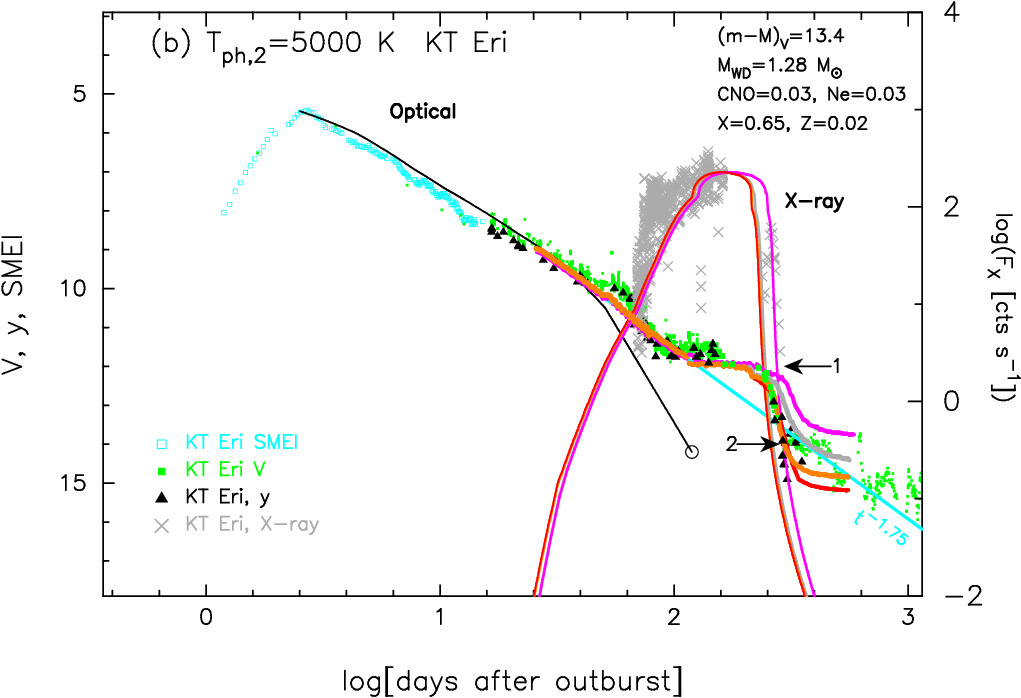}{0.75\textwidth}{}
          }
\caption{
Same as Figure \ref{kt_eri_only_v_x_big_disk_4500k_logscale}b, but
for $M_{\rm WD}= 1.28 ~M_\sun$ (Ne3).  The parameters of each model are
tabulated in Table \ref{2nd_plateau_parameters}.
(a) We assume the photospheric temperature of the companion star to be
$T_{\rm ph,2}=6,200$ K \citep{schaefer22wh}.
The black line depicts the FF+BB model light
curve for KT Eri.  The orange line reproduces both the first plateau
(1: $V\sim 12.0$ mag) and the second plateau (2: $V\sim 14.1$ mag), in which
the mass-accretion rate to the WD through the accretion disk is assumed
to be $\dot M_{\rm acc}= 1\times 10^{-9} ~M_\sun$ yr$^{-1}$ and
the disk shape parameters of $(\alpha, \beta)=(0.9, 0.01)$ in the
second plateau.  The magenta line denotes the case of
$\dot M_{\rm acc}= 1\times 10^{-7} ~M_\sun$ yr$^{-1}$ and
$(\alpha, \beta)= (0.9, 0.3)$, and the red line corresponds to 
$\dot M_{\rm acc}= 1\times 10^{-9} ~M_\sun$ yr$^{-1}$ and
$(\alpha, \beta)= (0.3, 0.01)$.
The corresponding X-ray light curves are plotted by the same color
lines as those of model $V$ light curves, but the orange and red lines are
overlapped.
(b) Same as those in panel (a) except $T_{\rm ph,2}=5,000$ K.
We added the gray line of $\dot{M}_{\rm acc}= 3\times 10^{-8} ~M_\sun$
yr$^{-1}$ and $(\alpha, \beta)= (0.9, 0.3)$. 
\label{kt_eri_only_v_x_big_logscale_no3_ab}}
\end{figure*}

\subsection{Irradiated disk and companion star}
\label{irradiated_disk_companion} 

We assume that the photosphere of the companion star just fills
its Roche lobe as in Figure \ref{kt_eri_config}.
We also assume that the disk surface is cylindrically
symmetric around the WD.  
As already explained in Appendix \ref{disk_not_disrupted},
a large disk represents the photosphere of the disk surface flow (disk winds),
the radial velocity of which is as large as $\sim 1000$ km s$^{-1}$
(see Section \ref{emission_line_profile})
because it is driven by the nova winds of velocities $v_{\rm wind}\sim
2000-3000$ km s$^{-1}$.  Therefore, the disk surface flow is dynamical
(much larger than the orbital velocity of $v_{\rm orb}\sim 100$ km s$^{-1}$)
and does not need to follow the gravitational potential
(or the Roche potential).  
It is reasonable that we assume axisymmetric flow when the
nova wind is axisymmetric (or spherically symmetric).

The size of the disk is defined by
\begin{equation}
R_{\rm disk}= \alpha R_{\rm RL,1},
\label{disk_radius_alpha}
\end{equation}
and the height of the disk at the edge is given by
\begin{equation}
H_{\rm disk}= \beta R_{\rm disk}.
\label{disk_height_beta}
\end{equation}
The surface height $z$ of the disk at the equatorial distance 
$\varpi= \sqrt{x^2+y^2}$ from the center of the WD is assumed to be
\begin{equation}
z = \left({{\varpi} \over {R_{\rm disk}}}\right) H_{\rm disk},
\label{disk_shape_winds}
\end{equation}
during the wind phase, but
\begin{equation}
z = \left( {{\varpi} \over {R_{\rm disk}}}\right)^2 H_{\rm disk},
\label{disk_shape_sss}
\end{equation}
after the winds stop.

\citet{hac03ka, hac03kb, hac03kc} proposed a large disk ($\alpha 
\sim 1.5-3.0$, $\beta=0.05$) during the wind phase to 
reproduce the temporal variations of the $V$ light curves of various
supersoft X-ray sources (SSSs), such as the recurrent nova CI Aql
\citep{hac03ka}, the Large Magellanic Cloud (LMC) SSS RX J0513.9-6951
\citep{hac03kb}, and V Sge \citep{hac03kc}.  

Assuming an accretion disk whose edge is flaring up
by the stream impact from the L1 point, \citet{sch97mm} explained
the orbital phase brightness variation of the supersoft X-ray sources
in the LMC.  We also assume that the disk edge is elevating up to 0.3
times the disk size, $\beta\le 0.3$ in Equation (\ref{disk_height_beta}),
after the winds stop, that is, in the SSS phase of KT Eri.

We adopt a disk whose size is 1.3 times
the Roche-lobe radius of the WD, as shown in Figure \ref{kt_eri_config}a,
during the nova wind phase.  The $\alpha = 1.3$ is taken from the 
observational result on the 2022 outburst of U Sco \citep[see Figure 5
of ][]{mura24ki}.  After the nova winds stop, we assume the size
of the disk to be 0.9 times the Roche-lobe radius of the WD,
$\alpha= 0.9$ in Equation (\ref{disk_radius_alpha}),
as shown in Figure \ref{kt_eri_config}b. 

The photospheric mesh surfaces in Figure \ref{kt_eri_config}, i.e., 
of the disk and companion star, are irradiated by the
central hot WD.  Then, the photospheric temperatures increase and emit
photons at these increased temperatures.
Such irradiation effects are all included in the
calculation of the $V$ light curve reproduction \citep[see][for the
partition of each surface and calculation method of irradiation]{hac01kb}.
Note that, in Figure \ref{kt_eri_config}, gas is optically thin
outside the mesh surfaces 
(i.e., photospheres of the disk, WD, and companion star).
In our irradiation calculation, we neglect absorption by optically thin
gas outside the mesh surfaces.

\subsection{Composite light curves: first plateau}
\label{temporal_variation_light_curve}

We have calculated the composite light curves as follows. 
In an early phase of the outburst, free-free emission 
from the wind dominates the $V$ band flux and
the FF+BB model light curve well follows the $V,y$ light curves
until day $\sim 40$.
When the WD photosphere shrinks to $R_{\rm ph}\sim 0.3 ~R_\sun$
($\sim 0.08 ~R_{\rm RL,1}$), i.e., $\sim 30$ days after the outburst,
the irradiated part of the disk had already emerged from the WD photosphere.
The large disk begins to contribute to the $V$ band luminosity 
comparably to the FF+BB flux. 

We divide the surfaces of the WD, disk, and companion star into small 
segments as in Figure \ref{kt_eri_config}.  Each patch of the WD emits
photons as a blackbody of the temperature $T_{\rm ph}$. 
Each patch of the disk and companion star absorbs photons from the WD,
heats up, and then emits blackbody photons with a higher temperature
than the original temperature.
We calculate its $V$ flux with a standard $V$ filter response
function and sum them up to obtain the total $V$ flux.
The calculation method is described in \citet{hac01kb}. 
Figure \ref{kt_eri_only_v_x_big_logscale_no3_ab} depicts our model $V$
light curves for such an irradiated disk (and companion star) as already
described by Equation (\ref{luminosity_summation_wd_disk_comp_v-band}) 
in Figure \ref{kt_eri_only_v_x_big_disk_4500k_logscale}b.

%

Our model well reproduces the brightness for the value of
$\alpha= 1.3$ (and $\beta=0.05$) during the wind phase.  The thickness
of $\beta=0.05$ is not strictly constrained but is likely to be
geometrically thin because its surface is blown in the wind.    

The $V,y$ brightnesses in the first plateau depends only on the size
and shape of the disk and does not depend on 
the mass-accretion rate, $\dot{M}_{\rm acc}$, or the photospheric
temperature of the companion star, $T_{\rm ph,2}$,
because the irradiation effect of the disk dominates the optical flux
in the SSS phase.  We need the size of the disk to be as large as
$\alpha \gtrsim 0.9$ (for $\beta=0.3$).  If we reduce the height
of the flaring edge down to $\beta=0.1$, we do not reproduce the brightness
of $V\sim 12.0$ in the first plateau.
Thus, we adopt a flaring edge of $\beta=0.3$.


\begin{deluxetable}{lllllll}
\tabletypesize{\scriptsize}
\tablecaption{Parameters and brightnesses of the second plateau
\label{2nd_plateau_parameters}}
\tablewidth{0pt}
\tablehead{
\colhead{$M_{\rm WD}$} & \colhead{$T_{\rm ph,2}$} & \colhead{$-\dot{M}_2$} 
& \colhead{$\alpha$} & \colhead{$\beta$} & \colhead{$V$\tablenotemark{a}}
& comment \\
($M_\sun$) & (~K~)  & ($M_\sun$ yr$^{-1}$) &  &  &  ( mag ) &
}
\startdata
1.28 & 6200 & $1\times 10^{-7}$ & 0.9 & 0.3 & 13.7 & \ref{kt_eri_only_v_x_big_logscale_no3_ab}a,magenta\tablenotemark{b}  \\
1.28 & 6200 & $1\times 10^{-7}$ & 0.9 & 0.01 & 13.7 & \\
1.28 & 6200 & $1\times 10^{-8}$ & 0.9 & 0.01 & 13.9 & \\
1.28 & 6200 & $1\times 10^{-9}$ & 0.9 & 0.01 & 14.0 & \ref{kt_eri_only_v_x_big_logscale_no3_ab}a,orange \\
1.28 & 6200 & $1\times 10^{-9}$ & 0.9 & 0.1 & 13.9  & \\
1.28 & 6200 & $1\times 10^{-9}$ & 0.9 & 0.3 & 13.7  & \\
1.28 & 6200 & $1\times 10^{-9}$ & 0.3 & 0.01 & 14.4 & \ref{kt_eri_only_v_x_big_logscale_no3_ab}a,red \\
1.28 & 6200 & $1\times 10^{-9}$ & 0.1 & 0.01 & 14.5 & \\
1.28 & 6200 & $1\times 10^{-10}$ & 0.9 & 0.01 & 14.0 & \\
\hline
1.28 & 5000 & $1\times 10^{-7}$ & 0.9 & 0.3 & 13.8 & \ref{kt_eri_only_v_x_big_logscale_no3_ab}b,magenta  \\
1.28 & 5000 & $1\times 10^{-7}$ & 0.9 & 0.01 & 14.4  & \\
1.28 & 5000 & $3\times 10^{-8}$ & 0.9 & 0.3 & 14.2 & \ref{kt_eri_only_v_x_big_logscale_no3_ab}b,gray  \\
1.28 & 5000 & $1\times 10^{-8}$ & 0.9 & 0.01 & 14.7  & \\
1.28 & 5000 & $1\times 10^{-9}$ & 0.9 & 0.01 & 14.8 & \ref{kt_eri_only_v_x_big_logscale_no3_ab}b,orange  \\
1.28 & 5000 & $1\times 10^{-9}$ & 0.9 & 0.1 & 14.6  & \\
1.28 & 5000 & $1\times 10^{-9}$ & 0.9 & 0.3 & 14.4  & \\
1.28 & 5000 & $1\times 10^{-9}$ & 0.3 & 0.01 & 15.2 & \ref{kt_eri_only_v_x_big_logscale_no3_ab}b,red  \\
1.28 & 5000 & $1\times 10^{-9}$ & 0.1 & 0.01 & 15.3  & \\
1.28 & 5000 & $1\times 10^{-10}$ & 0.9 & 0.01 & 14.8  & \\  
\hline
1.3 & 5000 & $2\times 10^{-7}$ & 0.9 & 0.3 & 14.0 & \\  
1.3 & 5000 & $1\times 10^{-9}$ & 0.9 & 0.01 & 14.8 & \\  
1.3 & 5000 & $1\times 10^{-9}$ & 0.3 & 0.01 & 15.2 &   \\
1.3 & 5000 & $1\times 10^{-11}$ & 0.01 & 0.01 & 15.3 &  \\
\hline
1.3 & 4500 & $2\times 10^{-7}$ & 0.9 & 0.3 & 14.1 & \ref{kt_eri_only_v_x_big_disk_4500k_logscale}b,magenta \\  
1.3 & 4500 & $1\times 10^{-7}$ & 0.9 & 0.01 & 14.6 & \ref{kt_eri_only_v_x_big_disk_4500k_logscale}b,orange \\  
1.3 & 4500 & $1\times 10^{-8}$ & 0.9 & 0.01 & 15.1 & \\  
1.3 & 4500 & $1\times 10^{-9}$ & 0.9 & 0.01 & 15.3 & \ref{kt_eri_only_v_x_big_disk_4500k_logscale}b,red  \\  
1.3 & 4500 & $1\times 10^{-9}$ & 0.3 & 0.01 & 15.6 &  \\
1.3 & 4500 & $1\times 10^{-11}$ & 0.01 & 0.01 & 15.8 & \ref{kt_eri_only_v_x_big_disk_4500k_logscale}b,gray    
\enddata
\tablenotetext{a}{
The model $V$ magnitude at the second plateau for the distance modulus
in the $V$ band of $(m-M)_V=13.4$.  The inclination angle is 
$i=40.9\arcdeg$ for $M_{\rm WD}=1.28 ~M_\sun$ (Ne3), the first 19 cases, 
but $i=41.2\arcdeg$ for $M_{\rm WD}=1.3 ~M_\sun$ (Ne3), the last 10 models.
}
\tablenotetext{b}{
Magenta line in Figure \ref{kt_eri_only_v_x_big_logscale_no3_ab}a. 
}
\end{deluxetable}

\subsection{Brightness in the second plateau}
\label{brightness_2nd_plateau}

Unlike the first plateau, the $V$ brightness in the second plateau 
depends on the mass-accretion rate.  
We calculate the $V$ magnitudes with a combination of 
parameters ($\dot {M}_2, \alpha, \beta$) for the second plateau.
The results are listed in Table \ref{2nd_plateau_parameters}. 
Here, the mass transfer rate from the companion is equal 
to the mass accretion rate onto the WD, $\dot {M}_2$ $=-\dot{M}_{\rm WD}$, 
under the mass conservation condition of $\dot{M}_{\rm WD}+\dot M_2=0$.
This means that, after the nova winds stop,
we neglect the effect of disk winds, L2 outflow, or
winds from the companion star.

Assuming that the mass-losing rate of the companion star is
$\dot M_2= -1.0\times 10^{-9} ~M_\sun$ yr$^{-1}$
and the disk shape of ($\alpha, \beta$) = (0.9, 0.01) in the second plateau,
we calculated the model $V$ light curve (orange line in Figure
\ref{kt_eri_only_v_x_big_logscale_no3_ab}a). 
The orange line reproduces 
the second plateau ($V\approx 14.0$).
The brightness in the second plateau is also tabulated in Table
\ref{2nd_plateau_parameters}.
The effect of impact of the L1 stream becomes weak for such a low
mass-transfer rate of $\dot M_2= -1.0\times 10^{-9} ~M_\sun$ yr$^{-1}$,
so that we adopt a geometrically thin accretion disk ($\beta=0.01$). 

After the second plateau, the nova enters the quiescent phase and 
the $V,y$ magnitudes go down to the quiescent level of $V\sim 15.1$. 
Our model (orange line)
in Figure \ref{kt_eri_only_v_x_big_logscale_no3_ab}a, however,  
does not decay down to $V\sim 15.1$, but keep staying at $V\approx 14.0$.
We changed the mass-losing rate $\dot{M}_2$ from $-1\times 10^{-7}$
to $-1\times 10^{-10} ~M_\sun$ yr$^{-1}$, but the brightness decreases
only from $V=13.7$ to 14.0 for ($\alpha, \beta$) = (0.9, 0.01), as
tabulated in Table \ref{2nd_plateau_parameters}.

Even if we change the size and shape of the disk, from $\alpha=0.9$ to 0.1,
from $\beta=0.3$ to 0.01, the $V$ magnitude varies between
13.7 and 14.5.  We concluded that the photospheric temperature
of the companion star, $T_{\rm ph,2}= 6,200$ K \citep{schaefer22wh},
is too high to be compatible with the observed $V$ brightness
in the quiescent phase ($V\sim 15.1 \pm 0.7$).

\subsection{Brightness in quiescence}
\label{brightness_quescent_phase}

We have to reduce the photospheric temperature of the companion star
from $T_{\rm ph,2}= 6,200$ K to $T_{\rm ph,2}= 5,000$ K,
and calculate the brightness in the quiescent phase.
The orange line in Figure \ref{kt_eri_only_v_x_big_logscale_no3_ab}b
shows a model of $\dot{M}_2= -1\times 10^{-9} ~M_\sun$ yr$^{-1}$,
$\alpha=0.9$, and $\beta=0.01$, which reproduces well the brightness
of $V\sim 14.8$ in the quiescent phase.

However, we do not reproduce the lowest brightness of $V\sim 15.8$
by this $T_{\rm ph,2}= 5,000$ K model,
as tabulated in Table \ref{2nd_plateau_parameters}.
We must further decrease the photospheric temperature of the companion
star down to $T_{\rm ph,2}= 4,500$ K.
This will be examined in Section \ref{1.3Msun_white_dwarf}.

\subsection{Duration of the hydrogen burning}
\label{duration_hydrogen_burning}

Figure \ref{kt_eri_only_v_x_big_disk_4500k_logscale}a shows that
the Swift X-ray count rate quickly dropped on day $\sim 280$, and 
Figure \ref{vflux_o3_he2}b shows that the intensity of \ion{He}{2}
4686\AA\   line started to drop on day $\sim 240$.
We regard that these epochs correspond to the end of hydrogen burning.
Figure \ref{all_mass_kt_eri_x55z02o10ne03} shows, however, that all of
our best-fit optical models begin 
to decay earlier than the X-ray count rate does.
 Because we have already fixed two model parameters, i.e., the WD mass
and chemical composition of the envelope, and the X-ray turnoff time
does not depend on the disk parameters, there remains one parameter,
the mass-accretion rate to the WD, which could extend the period of
hydrogen burning.  We adopt four mass-accretion rates, 
$\dot{M}_{\rm acc}= 1\times 10^{-10}$, 
$1\times 10^{-9}$, $1\times 10^{-8}$, and
$1\times 10^{-7} ~M_\sun$ yr$^{-1}$, and examine how the mass-accretion
rate changes the duration of hydrogen burning, i.e., the period of the
SSS phase.  Especially, high mass-accretion rates supply fuel effectively
to hydrogen burning and make its duration significantly longer, as
expected from Equation (\ref{nova_evoluion_eq}).
These parameters are summarized in Table \ref{2nd_plateau_parameters}.

Among these four $\dot{M}_{\rm acc}$ models,
the $\dot{M}_{\rm acc}= 1\times 10^{-7} ~M_\sun$
yr$^{-1}$ model catches up the X-ray decay on day $\sim 280$, as indicated by
the thin magenta lines in Figure \ref{kt_eri_only_v_x_big_logscale_no3_ab}a
and b.  On the other hand, this model (the thick magenta lines in both
Figure \ref{kt_eri_only_v_x_big_logscale_no3_ab}a and b)
does not correctly follow the optical $V,y$ observation;
the decay to the second plateau is too late.

In this way, we have examined $1.28 ~M_\sun$ WD (Ne3) models
with various parameters, but there seem to be no models that satisfy
all the observational requirements.

\subsection{The $1.3 ~M_\sun$ WD (Ne3)}
\label{1.3Msun_white_dwarf}

We finally adopt a $1.3 ~M_\sun$ WD (Ne3) for our KT Eri model.
We first assume that $T_{\rm ph,2}= 5,000$ K for $M_2= 1.0 ~M_\sun$.
The model of $(-\dot{M}_2, ~\alpha, ~\beta)
=(2\times 10^{-7} ~M_\sun$ yr$^{-1}, ~0.9, ~0.3)$
reproduces both the first ($V\sim 12.1$) and
second ($V\sim 14.0$) plateau
including their temporal variations as well as their $V$ brightnesses. 
This means that the mass-accretion rate and the size and shape of
the disk are kept constant in the first and second plateaus
until the second plateau ends.

The cases of $(-\dot{M}_2, ~\alpha, ~\beta) =(1\times 10^{-9} ~M_\sun$
yr$^{-1}, ~0.9, ~0.01)$, $(1\times 10^{-9} ~M_\sun$ yr$^{-1}, ~0.3, ~0.01)$,
and $(1\times 10^{-11} ~M_\sun$ yr$^{-1}, ~0.01, ~0.01)$ show
the brightness of $V\sim 14.8$, 15.2, and 15.3, respectively, 
in quiescence, as summarized in Table \ref{2nd_plateau_parameters}.
Thus, the average brightness $V\sim 15.1$ in quiescence
can be explained by this $M_{\rm WD}= 1.3 ~M_\sun$ WD and
$T_{\rm ph,2}= 5,000$ K model.

Therefore, we may conclude that
the transition from the second plateau to the quiescence is driven
by the reductions of the mass-transfer rate from $-\dot{M}_2=
2\times 10^{-7}$ to $1\times 10^{-9} ~M_\sun$ yr$^{-1}$
and the flaring-up factor from $\beta=0.3$ to $0.01$.
This reminds us that high mass transfer rates such as
$2\times 10^{-7} ~M_\sun$ yr$^{-1}$ makes a spray at the edge of the disk
as illustrated by \citet{sch97mm}.

The faintest brightness in quiescence, however, is as low as 
about $V\sim 15.8$ \citep{schaefer22wh}.  This low brightness is
obtained for $T_{\rm ph,2} = 4,500$ K, as shown by the gray line
in Figure \ref{kt_eri_only_v_x_big_disk_4500k_logscale}b and Table
\ref{2nd_plateau_parameters}.
In this case, the parameters are $(-\dot{M}_2, ~\alpha, ~\beta)
=(1\times 10^{-11} ~M_\sun$ yr$^{-1}, ~0.01, ~0.01)$.  This means that
the WD and disk component are negligible.  In other words, the mass-transfer
virtually stops and then the disk almost disappears. 
If we could obtain spectral energy distribution of KT Eri
in the faintest state,
we would determine the nature of the companion star. 
Our estimate on the variable $-\dot{M}_2$ is consistent with 
\citet{sun20od}'s low mass-accretion rate of $\dot{M}_{\rm acc} \approx
1.9 \times 10^{-10} ~(d/3.7{\rm ~kpc})^2 
~M_\sun$ yr$^{-1}$ about 8.3 yr after the outburst.
The origin of such a variation in the mass-transfer rate is unclear.  

\citet{schaefer22wh} obtained a very high mass-losing rate of
$\dot{M}_2= -3.5\times 10^{-7} ~(d/5.1{\rm ~kpc})^2
~M_\sun$ yr$^{-1}$ as an average in
quiescence.  If this is the case, hydrogen burning never stops
($-\dot{M}_2 \gtrsim \dot{M}_{\rm steady}\approx 3\times 10^{-7} ~M_\sun$
yr$^{-1}$ for $M_{\rm WD}= 1.3 ~M_\sun$), the SSS phase never ends, 
the  $V,y$ brightnesses keep $V,y \sim 12.0$ for a long time, and never
drops to $V\sim 15$.

We may conclude that the average mass-transfer rate is as low as 
$\dot{M}_2 \sim -1\times 10^{-9} ~M_\sun$ yr$^{-1}$ ($V\sim 15.3$
and $\dot{M}_{\rm WD} \ll \dot{M}_{\rm steady}$) before the outburst and
increases to $-\dot{M}_2 \sim 2\times 10^{-7} M_\sun$ yr$^{-1}$
when the winds stopped on day $\sim 100$.  It should be noted that,
in this model, we can explain both the brightness in the second plateau
and the SSS duration at the same time with a single mass accretion rate
of $\dot{M}_{\rm acc}=2\times 10^{-7} ~M_\sun$ yr$^{-1}$.

Finally, we have checked the case of $M_{\rm WD}= 1.32 ~M_\sun$ (Ne3),
because the three cases of $M_{\rm WD}= 1.28$, 1.3, and $1.32 ~M_\sun$
(Ne3) satisfy the observational constraints except for the decay of
supersoft X-ray light curve in Figure \ref{all_mass_kt_eri_x55z02o10ne03}b
(see Appendix \ref{opticall_thick_wind_model}).
We adopt $\dot{M}_2 \sim -2\times 10^{-7} ~M_\sun$ yr$^{-1}$
and $(\alpha , \beta)=(1.3, 0.05)$ in the wind phase, $(0.9, 0.3)$
in the SSS phase (the first plateau) and in the second plateau.
The $V$ brightness reproduces $V\sim 12.0$ in the first plateau and
$V\sim 14.1$ in the second plateau but each start and end epochs are
too early to be compatible with the observation.  For example,
the end epoch of hydrogen burning is day $\sim 190$, about 50 days earlier.
Thus, we may exclude the WD masses of 1.32 $M_\sun$ and
conclude that the WD mass is $1.28 ~M_\sun < M_{\rm WD} < 1.32 ~M_\sun$.

\section{Distance moduli in the $B$, $V$, and $I_{\rm C}$ bands}
\label{distance_moduli_bvi}


\begin{figure*}
\gridline{\fig{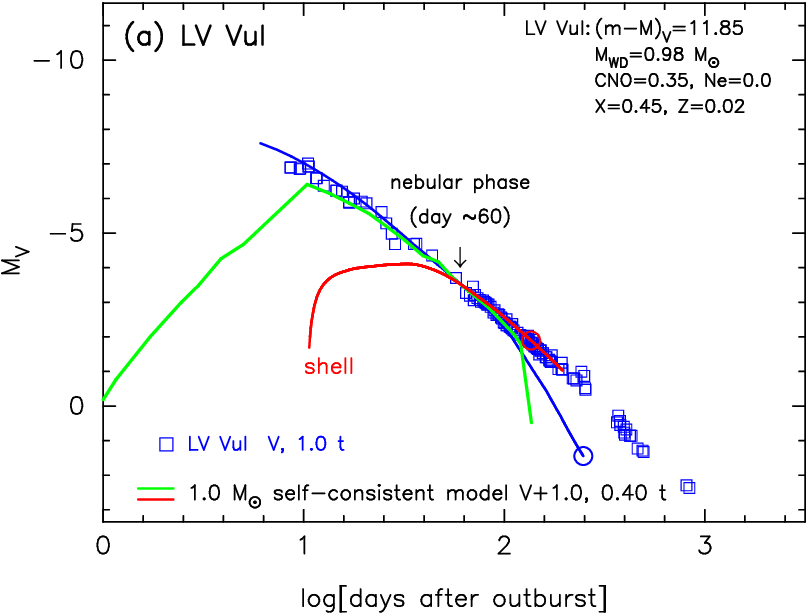}{0.51\textwidth}{}
          \fig{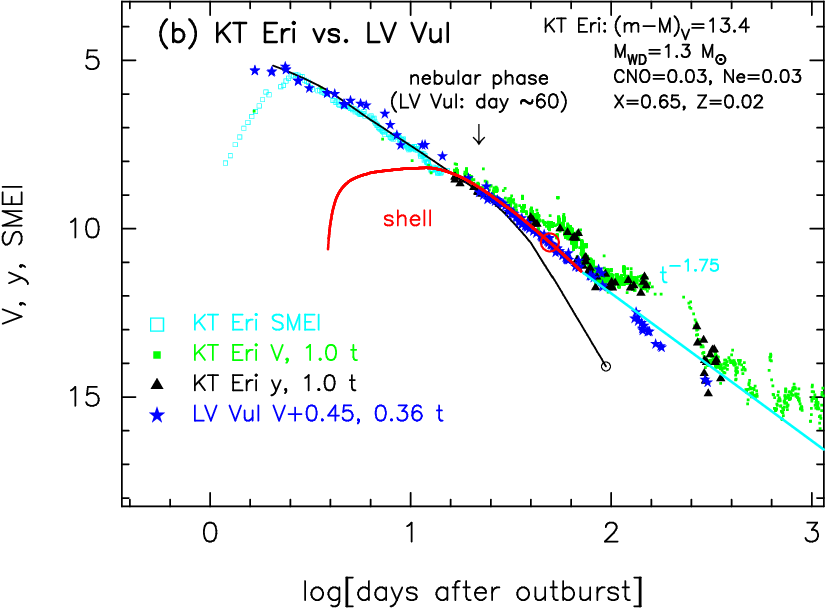}{0.51\textwidth}{}
          }
\gridline{\fig{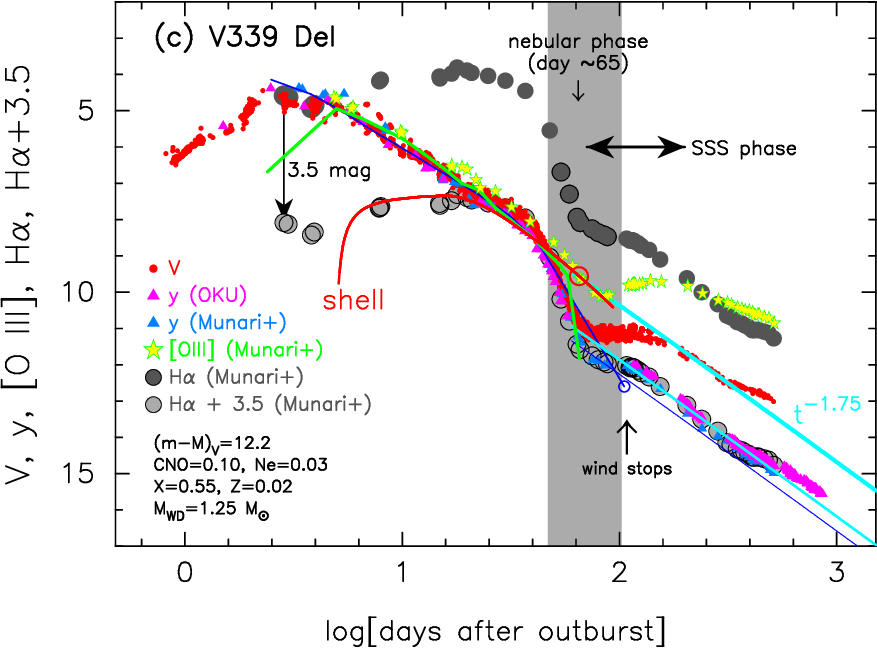}{0.51\textwidth}{}
          \fig{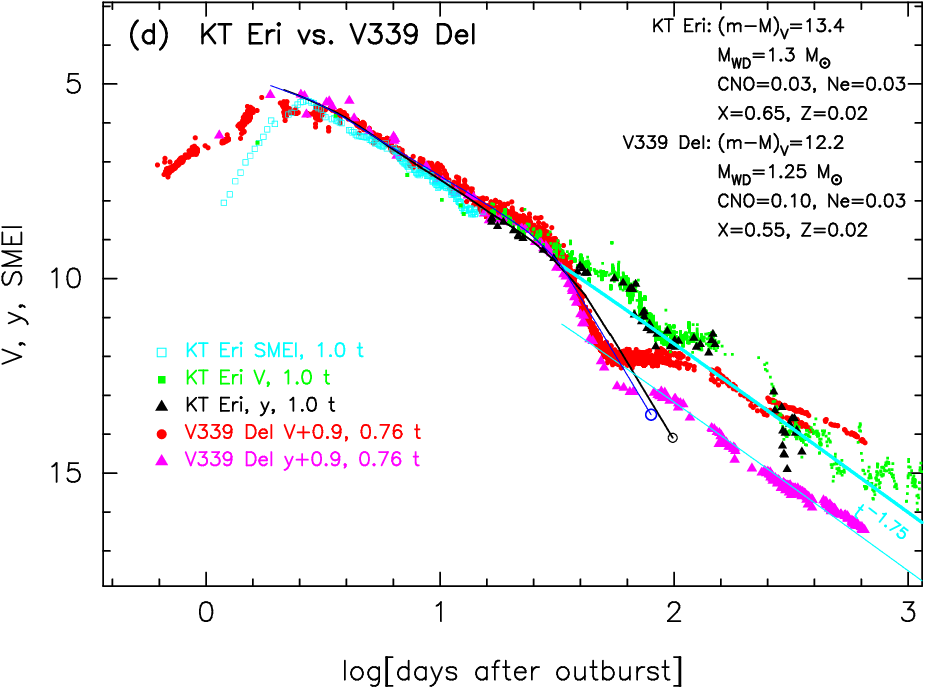}{0.51\textwidth}{}
          }
\caption{
(a) The $V$ light curve of the classical nova LV Vul
for $(m-M)_V= 11.85$ \citep{hac21k}.
The $V$ data are the same as those in \citet{hac16kb}. 
We indicate the start of the nebular phase of LV Vul about 60 days
after the outburst \citep{hac16kb}.
The blue line depicts a best-fit FF+BB model in \citet{hac19k}, that is,
a $0.98 ~M_\sun$ WD (CO3: 
$X=0.45$, $Y=0.18$, $Z=0.02$, $X_{\rm CNO}=0.35$, $X_{\rm Ne}=0$)
among \citet{kat94h}'s nova models.  
The green one denotes the free-free plus blackbody (FF+BB) flux
\citep{kat22sha, hac23k} and
the red one does the emission from the shocked shell based on the
shock model of \citet{hac22k}.  The open red circle on
the red line indicates the end epoch of winds \citep{hac23k}. 
These two lines are calculated from the time-stretched
1.0 $M_\sun$ self-consistent nova outburst model \citep{hac23k}:
$f_{\rm s} = 0.4$ and $\Delta V = +1.0$ in Equation
(\ref{overlap_brigheness}) for target= LV Vul and template= 1.0 $M_\sun$
WD model.
(b) Comparison between KT Eri and LV Vul with $f_{\rm s}= 0.36$ and
$\Delta V= +0.45$.
The black line denotes our FF+BB model light curve of a $1.3 ~M_\sun$ WD
(Ne3) for KT Eri.
We indicate the slope of $L_V\propto t^{-1.75}$ (the thick cyan line).
(c) The $V$, $y$, [\ion{O}{3}], H$\alpha$, 
and H$\alpha + 3.5$ mag light curves of the classical nova V339 Del.
The data are the same as those in Figure 8(a) of \citet{hac24km}.
The gray-shaded region represents the substantial absorption phase
by a dust shell.  We indicate the SSS
phase by the two-headed arrow labeled ``SSS phase.''  The curved blue line
is the FF+BB light curve of a 1.25 $M_\sun$ WD (Ne2) model for V339 Del,
which is taken from \citet{hac24km}. 
The green and red lines are the same as those in panel (a),
but with $f_{\rm s}= 0.145$ and $\Delta V= +1.45$.
The H$\alpha + 3.5$ mag light curve first follows the shape of the shell
emission because the H$\alpha$ line comes from the shocked shell
\citep{hac24km}.
(d) Comparison between KT Eri and V339 Del with the time-stretching factor
of $f_{\rm s}= 0.76$ and $\Delta V= +0.9$.
The other symbols and lines are the same as those in panels (b) and (c).
\label{kt_eri_v339_del_lv_vul_v_small_logscale_no3}}
\end{figure*}

\subsection{The time-stretching method}
\label{time-stretching_method}

The distance to KT Eri is a little bit controversial between
$d=4.2$ kpc \citep[Gaia eDR3,][the present work]{schaefer22b}
and $d=5.1$ kpc \citep{schaefer22wh}, as introduced in Section
\ref{distance_reddening}.
In this Appendix, we apply the time-stretching method to KT Eri and
LV Vul, and obtain the distance to the nova \citep{hac10k, hac15k, hac16kb}.

This method is based on the similarity of nova light curves.
Adopting appropriate time-stretching parameters, we are able to overlap
two nova light curves even if the two nova speed classes are different.
If the two nova $V$ light curves, i.e.,
one is called the template and the other is called the target,
$(m[t])_{V,\rm target}$ and $(m[t])_{V,\rm template}$
overlap each other after time-stretching of a factor of $f_{\rm s}$
in the horizontal direction and shifting vertically down by $\Delta V$, i.e.,
\begin{equation}
(m[t])_{V,\rm target} = \left((m[t \times f_{\rm s}])_V
+ \Delta V\right)_{\rm template},
\label{overlap_brigheness}
\end{equation}
their distance moduli in the $V$ band satisfy
\begin{eqnarray}
(m&-&M)_{V,\rm target} \cr
&=& \left( (m-M)_V + \Delta V\right)_{\rm template} - 2.5 \log f_{\rm s},
\label{distance_modulus_formula}
\end{eqnarray}
where $m_V$ and $M_V$ are the apparent and absolute $V$ magnitudes,
and $(m-M)_{V, \rm target}$ and $(m-M)_{V, \rm template}$ are
the distance moduli in the $V$ band
of the target and template novae \citep[e.g.,][]{hac20skhs}.

Figure \ref{kt_eri_v339_del_lv_vul_v_small_logscale_no3}a shows 
the $V$ light curve of the classical nova LV Vul.  Here, we time-stretch
the light curves of \citet{kat22sha}'s self-consistent nova model
(green and red lines) against that of LV Vul (open blue squares)
by the time-shift of $\Delta \log t = \log f_{\rm s} = \log 0.4 = -0.4$
in the horizontal direction and the vertical shift by $\Delta V = +1.0$.
The green line denotes the free-free plus blackbody (FF+BB) model light curve
while the red line represents the shell emission of the self-consistent
nova model, both of which were calculated by \citet{hac23k}.
The shell emission comes from the shocked shell far outside the nova
photosphere (see Figure \ref{kt_eri_line_profile}a and c).
In Figure \ref{kt_eri_v339_del_lv_vul_v_small_logscale_no3}b,
we further time-stretch the $V$ light curve of LV Vul and overlap
it with the $V,y$ light curves of KT Eri.
The time-stretching factor is $f_{\rm s}=0.36$, that is, the time-shift by
$\Delta \log t = \log f_{\rm s} = -0.44$ in the horizontal direction
and the vertical shift by $\Delta V = +0.45$ mag.
The $V$ light curve (filled blue stars) declines almost
along with the same universal decline law of $L_V \propto t^{-1.75}$
(thick cyan line) over the nebular phase.

Applying Equation (\ref{distance_modulus_formula}) to KT Eri and LV Vul in
Figure \ref{kt_eri_v339_del_lv_vul_v_small_logscale_no3}b, we obtain
\begin{eqnarray}
(m&-&M)_{V, \rm KT~Eri} \cr
&=& ((m - M)_V + \Delta V)_{\rm LV~Vul} - 2.5 \log 0.36 \cr
&=& 11.85 + 0.45\pm0.2 + 1.1 = 13.4\pm0.2,
\label{distance_modulus_v_kt_eri_lv_vul}
\end{eqnarray}
where we adopt $(m - M)_{V, \rm LV ~Vul}= 11.85$ from \citet{hac21k}.
Thus, we obtain $(m-M)_{V, \rm KT~Eri}= 13.4\pm0.2$, which is 
consistent with the Gaia eDR3 distance of $d=4.1^{+0.5}_{-0.4}$ kpc
\citep{bai21rf} and \citet{schaefer22b}'s distance of
$d=4211^{+466}_{-296}$ pc
together with the extinction of $E(B-V)=0.08$
\citep{rag09bs}.
The distance modulus in the $V$ band of $(m-M)_{V, \rm KT~Eri}= 13.4\pm0.2$
is also supported by our $1.3 ~M_\sun$ WD model in Figure
\ref{kt_eri_only_v_x_big_disk_4500k_logscale}b.


\subsection{Universal decline law in the nebular phase}
\label{nova_nebular_phase}

Many novae show the universal decline law of $L_V\propto t^{-1.75}$ even
in the nebular phase \citep{hac06kb, hac10k, hac23k}.  On the other hand,
the FF+BB model light curves follow the universal decline law 
until the start of the nebular phase but, after that, depart from
the $V$ light curve (see, e.g., Figure 
\ref{kt_eri_v339_del_lv_vul_v_small_logscale_no3}a for LV Vul).
We discuss this issue for LV Vul and KT Eri based on the results of 
\citet{hac23k}.  The FF+BB light curve of the self-consistent nova model
(green line in Figure \ref{kt_eri_v339_del_lv_vul_v_small_logscale_no3}a)
reaches its maximum light and then declines almost along the universal
decline law of $L_V \propto t^{-1.75}$.  It rapidly decreases just before
optically thick winds stop (at the open red circle on the red line in
Figure \ref{kt_eri_v339_del_lv_vul_v_small_logscale_no3}a).  This is
because the free-free flux depends on the square of wind mass-loss rate,
i.e., $L_{V, \rm ff,wind}\propto \dot{M}_{\rm wind}^2$ from Equation
(\ref{free-free_flux_v-band}), and $\dot{M}_{\rm wind}$ is 
rapidly decreasing just before winds stop. 

Then, the density of winds is also rapidly decreasing and the photospheric
temperature increases \citep[see, e.g., Figure 5 of][for the temporal
variation of the photospheric temperature]{hac24km}.  Both the quick
decrease in the wind density and the rapid increase in the temperature
of the radiation field are suitable for the excitation of
[\ion{O}{3}] 4959, 5007\AA\  lines.
\citet{hac24km} analyzed the temporal flux variation of [\ion{O}{3}] lines
for V339 Del (Figure \ref{kt_eri_v339_del_lv_vul_v_small_logscale_no3}c)
and found that the [\ion{O}{3}] emission lines comes from
the shocked shell.  \citet{ara13ii} analyzed [\ion{O}{3}] 4959, 5007\AA\ 
lines of KT Eri obtained on day 294 (in the nebular phase).  Their 
decomposed doublet [\ion{O}{3}] line profiles show a broad 2,200 km s$^{-1}$
rectangular (pedestal) shape in the second plot of their Figure 2.
This is consistent with our interpretation that 
[\ion{O}{3}] 4959, 5007\AA\  lines come from the shocked shell
as illustrated in Figure \ref{kt_eri_line_profile}c.

The shell emission of the self-consistent nova model 
(red line in Figure \ref{kt_eri_v339_del_lv_vul_v_small_logscale_no3}a)
is calculated from the flux from the shocked shell (see Figure
\ref{kt_eri_line_profile}a and c) and eventually
dominates the $V$ luminosity in the nebular phase.
See Equation (B6) of \citet{hac23k} for the calculation of the red line.
In this LV Vul case in Figure 
\ref{kt_eri_v339_del_lv_vul_v_small_logscale_no3}a,
the shell emission in the $V$ band is dominated by [\ion{O}{3}] lines.
The fluxes of [\ion{O}{3}] 4959, 5007\AA\  lines
follows the same universal decline law of $L_V \propto t^{-1.75}$
in the nebular phase.
Thus, the shell emission takes over the FF+BB flux and 
extends our universal decline law of $L_V \propto t^{-1.75}$ 
even to the nebular phase (the red line in Figure 
\ref{kt_eri_v339_del_lv_vul_v_small_logscale_no3}b).

To summarize, the FF+BB flux decays rapidly after the nebular
phase started.  The flux of strong [\ion{O}{3}] emission lines
increases and fills the gap between the FF+BB flux and the $V$ observation.
This is the reason why many novae follow the universal decline law even
in the nebular phase.

If the dominant source in the $V$ band is [\ion{O}{3}] 4959, 5007\AA\  lines,
the $V$ magnitude should be much brighter than the $y$ magnitude in the
nebular phase, because the $y$ band filter is designed to avoid the
[\ion{O}{3}] 4959, 5007\AA\  lines.  Such examples are V1500 Cyg, V1668 Cyg,
and V339 Del, as already introduced in Section \ref{Vy_light_curve_same}.

We may conclude that the nebular phase of a nova is driven by
(1) the quick decrease in the wind mass-loss rate $\dot{M}_{\rm wind}$
and (2) the concurrent rapid rise of the photospheric temperature of a nova
$T_{\rm ph}$, both of which are inherent properties of a nova outburst.
We add (3) that [\ion{O}{3}] 4959, 5007\AA\  lines are mainly excited
in the shocked shell, which is far outside the photosphere.
This is because the shocked shell eventually collects about 90\% of
the ejected mass \citep{hac22k}.
Therefore, the line profiles of [\ion{O}{3}] 4959, 5007\AA\  are basically
rectangles (or pedestals) if the shocked shell is spherically
symmetric.


\begin{figure*}
\gridline{\fig{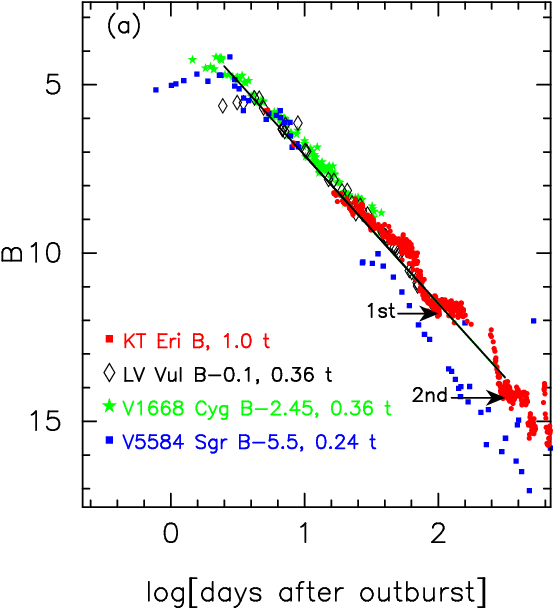}{0.475\textwidth}{}
          \fig{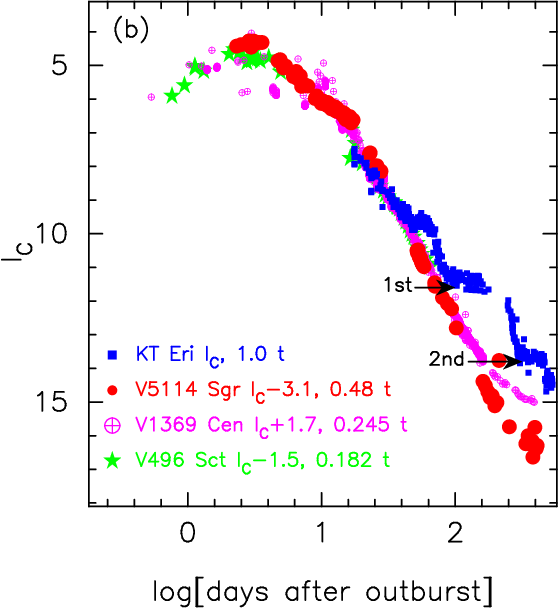}{0.475\textwidth}{}
          }
\caption{
(a) The $B$ light curve of KT~Eri as well as
those of LV~Vul, V1668~Cyg, and V5584~Sgr.  
The $B$ data of KT Eri are taken from VSOLJ, AAVSO, SMARTS, and \citet{ima12t}.
The $B$ data of the other novae are the same as those in \citet{hac19k,hac21k}.
Each light curve is horizontally moved by $\Delta \log t = \log f_{\rm s}$
and vertically shifted by $\Delta B$ with respect to that of KT Eri,
as indicated by ``LV Vul B$-$0.05, 0.36 t,'' which means
$\Delta B=-0.05$ and $\Delta \log t= \log f_{\rm s}= \log 0.36 = -0.44$.
The solid black line denotes the slope of $L_B \propto t^{-1.75}$,
where $L_B$ is the $B$ band luminosity.  This slope is the trend of
the universal decline law \citep{hac06kb}.  Horizontal arrows indicate
the ``1st'' and ``2nd'' plateaus corresponding to the first and second
plateaus in the $V,y$ magnitudes of KT Eri.  
(b) Same as panel (a), but for the $I_{\rm C}$ light curve.
The $I_{\rm C}$ data of KT Eri are taken from VSOLJ, AAVSO, and SMARTS. 
We add the $I_{\rm C}$ light curves of V5114 Sgr, V1369~Cen, and V496~Sct.
The $I_{\rm C}$ data of the other novae are the same as those
in \citet{hac21k}.
\label{distance_reddening_kt_eri_bvi_xxxxxx}}
\end{figure*}


\begin{figure}
\epsscale{1.0}
\plotone{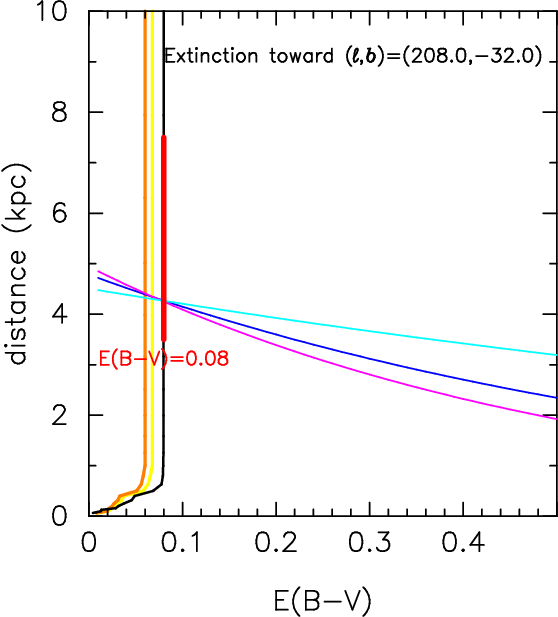}
\caption{
Various distance-reddening relations toward KT Eri.
The three thin lines of magenta, blue, and cyan
denote the distance-reddening relations given by 
$(m-M)_B= 13.47$, $(m-M)_V= 13.4$, and $(m-M)_I= 13.27$, respectively.
These three lines cross broadly at $d=4.2$ kpc and $E(B-V)=0.08$
(thick vertical red line).
Three thick black, orange, and yellow lines indicate the galactic
distance-reddening relation toward the direction of KT Eri
given by \citet{gre15}, \citet{gre18}, and \citet{gre19},
respectively. 
\label{distance_reddening_kt_eri_bvi_cross}}
\end{figure}

\subsection{Comparison with V339 Del}
\label{comparison_v339_del}

Figure \ref{kt_eri_v339_del_lv_vul_v_small_logscale_no3}c depicts the
light curves of the classical nova V339 Del in various bands
\citep[taken from Figure 8(a) of ][]{hac24km}, and 
Figure \ref{kt_eri_v339_del_lv_vul_v_small_logscale_no3}d displays the
light curves both of V339 Del and KT Eri.
V339 Del is a classical nova having a similar decay timescale to KT Eri.
V339 Del also has a flat plateau in the $V$ light curve
during the SSS phase \citep[e.g., Figure 1 of ][]{sho16ms} as denoted 
by the two-headed arrow labeled ``SSS phase'' 
in Figure \ref{kt_eri_v339_del_lv_vul_v_small_logscale_no3}c.
The $y$ magnitudes were well obtained from the optical peak to the very
late phase of the outburst \citep{mun15mm, hac24km}, as shown in Figure
\ref{kt_eri_v339_del_lv_vul_v_small_logscale_no3}c and d.
Unlike KT Eri, V339 Del shows a different property between the $V$ and
$y$ magnitudes:  the $V$ light curve shows a flat plateau, but the $y$ 
light curve does not.  In this subsection, we deeply compare V339 Del
with KT Eri and clarify the physical nature of the plateau phase in
KT Eri.


\begin{figure*}
\gridline{\fig{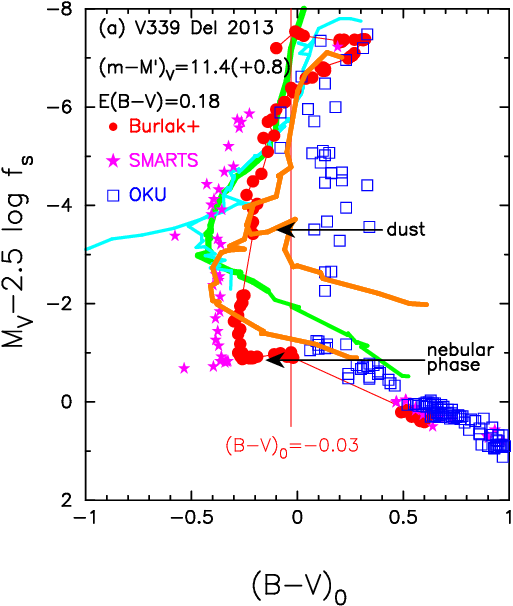}{0.42\textwidth}{}
          \fig{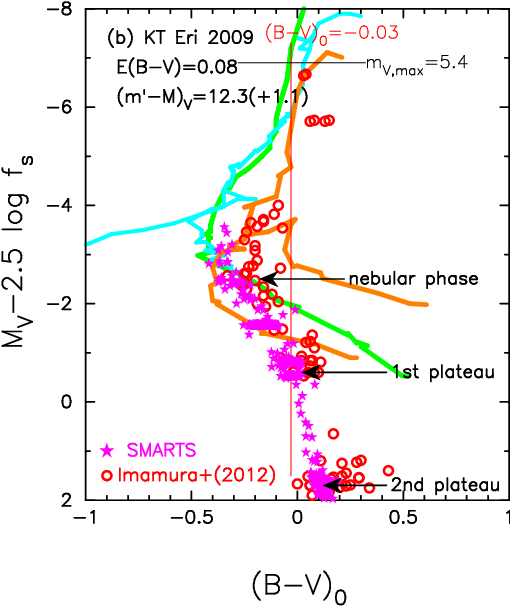}{0.42\textwidth}{}
          }
\gridline{\fig{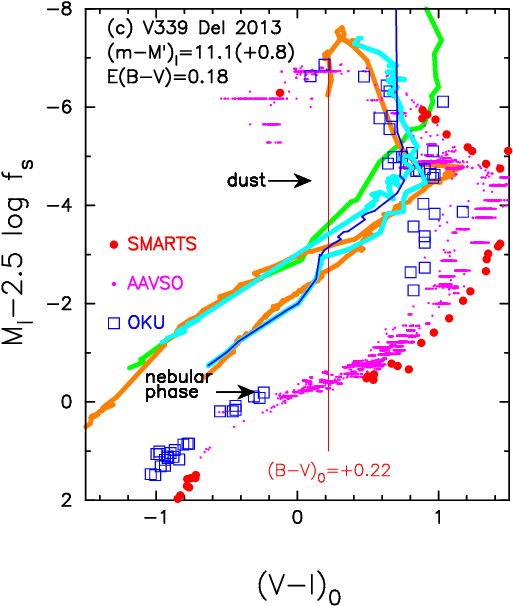}{0.42\textwidth}{}
          \fig{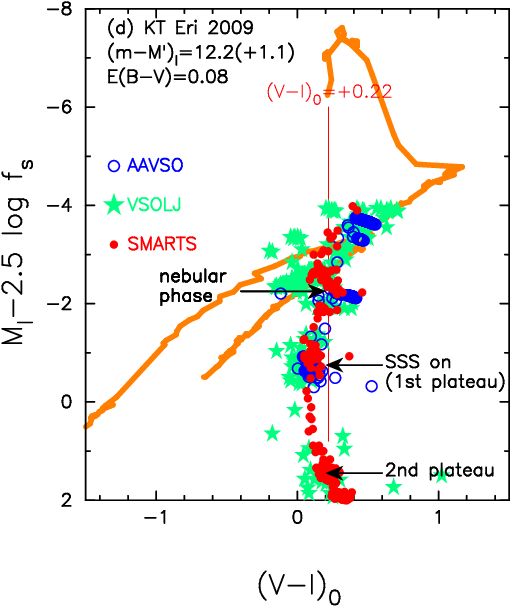}{0.42\textwidth}{}
          }
\caption{
(a) The time-stretched $(B-V)_0$-$(M_V-2.5 \log f_{\rm s})$ color-magnitude
diagram for V339~Del.  The data of V339~Del are taken from \citet{bur15a},
SMARTS \citep{wal12bt}, and OKU \citep{hac24km}.
The epochs of dust shell formation and
start of the nebular phase are indicated by arrows. 
The vertical solid red line of $(B-V)_0= -0.03$ is the
intrinsic color of optically thick free-free emission \citep{hac14k}.
The solid green, cyan, and orange lines denote the template tracks
of V1500~Cyg, V1974~Cyg, and LV~Vul, respectively.
The track of V1974~Cyg/LV~Vul splits into two branches in the later
phase.  See \citet{hac16kb} and \citet{hac19k}
for details of these template tracks.
(b) Same as in panel (a), but for KT Eri.
The data of KT~Eri are taken from SMARTS \citep{wal12bt} and \citet{ima12t}.
We specify the onset of nebular phase by a turning point of $(B-V)_0$ color,
though not a usual definition.
(c) The time-stretched $(V-I)_0$-$(M_I-2.5 \log f_{\rm s})$ color-magnitude
diagram for V339~Del.  The data of V339~Del are taken from AAVSO,
SMARTS \citep{wal12bt}, and OKU \citep{hac24km}.  The thick solid lines
of orange, green, blue, and cyan correspond to the outburst tracks of
LV Vul, V1500 Cyg, V382 Vel, and LV Vul (V496 Sct, V959 Mon),
which are taken from \citet{hac21k}. 
(d) Same as those in panel (c), but for KT Eri.
\label{hr_diagram_kt_eri_v339_del_bv_mfs}
}
\end{figure*}

Figure \ref{kt_eri_v339_del_lv_vul_v_small_logscale_no3}c shows five
light curves of $V$, $y$, [\ion{O}{3}], H$\alpha$, and H$\alpha + 3.5$
magnitudes of V339 Del.  The $V$ and $y$ light curves decline very
similarly until the nebular phase started on day $\sim 65$ \citep{mun13hdc}.
After day $\sim 65$, the $y$ magnitude decline along the thin blue line
of $L_y\propto t^{-1.75}$, whereas the $V$ magnitude levels off
for a while between day $\sim 65$ and day $\sim 140$, showing a flat
plateau phase, and then declines along the thick cyan line of
$L_V\propto t^{-1.75}$.

\citet{hac24km} presented a detailed light curve model of V339 Del and
clarified these complicated behavior.  The FF+BB light curve (Equation 
(\ref{luminosity_summation_flux_v-band})) of a $1.25 ~M_\sun$ WD (Ne2)
broadly reproduced the $V,y$ light curve until day $\sim 50$.
The H$\alpha+3.5$ magnitude follows the emission from the shocked shell 
in the early phase (until day $\sim 45$).
Thus, the H$\alpha$ is shock-origin, at least, in an early phase of V339 Del.
After day $\sim 45$, both the $V$ and $y$ light curve drastically drop 
along the FF+BB light curve because the model wind mass-loss rate
quickly drops and the model free-free flux from the nova wind sharply
drops \citep[$L_{V, \rm ff,wind}\propto \dot{M}_{\rm wind}^2$ from Equation
(\ref{free-free_flux_v-band}),][]{hac06kb}.
A part of the $V,y$ fluxes are absorbed by dust
\citep{hac24km}.  After day $\sim 65$, the $y$ light curve is monotonically
declining with the slope of $L_y\propto t^{-1.75}$ (thin blue or cyan line), 
which means the continuum emission declines as $L_y\propto t^{-1.75}$. 
On the other hand, the $V$ light curve levels off and keeps
a flat plateau of $V \sim 11$ between day $\sim 65$ and day $\sim 140$,
and then starts to decline with the same slope of $L_V\propto t^{-1.75}$
(thick cyan line).  Thus, the $V$ magnitude is about 1.5 mag 
brighter than the $y$ magnitude after day $\sim 140$.  
The [\ion{O}{3}] emission line fluxes contribute to the $V$ band flux
and fills the gap between $V$ and $y$ \citep{mun15mm}. 

A dust shell could be formed within a cool and dense shell
behind the radiative shock in a nova ejecta \citep[e.g.,][]{der17ml}.
The [\ion{O}{3}] emission line originates from the layer slightly outside
the dust shell in the shocked shell, because the [\ion{O}{3}] light curve
does not show a dust dip between day $\sim 45$ and day $\sim 140$
in Figure \ref{kt_eri_v339_del_lv_vul_v_small_logscale_no3}c.  This means
that each layer is ordered, the shock, dust, and [\ion{O}{3}] emission,
from inside to outside.  \citet{hac24km} concluded that the [\ion{O}{3}]
emission line comes from the outermost layer of the shocked shell.
The H$\alpha ~(+ 3.5$ mag), which shows a rapid decay after day $\sim 45$
compared with the shell emission light curve (red line),
resides at the shock and dust, so that it could be
suppressed and absorbed by dust (and cool environment).
In this way, the difference among $V$, $y$, and [\ion{O}{3}] tells us
how the shocked shell contributes to the luminosity.

The $V$ light curve of V339 Del shows a plateau during its SSS phase (from
day $\sim 72$ to day $\sim 200$) while the $y$ light curve does not 
(Figure \ref{kt_eri_v339_del_lv_vul_v_small_logscale_no3}c).
The plateau of V339 Del in the $V$ light curve is caused by the increase
in the [\ion{O}{3}] line emission, therefore, it does not mean
an irradiation effect of the disk.  Moreover,
V339 Del is a close binary having a short orbital period of
$P_{\rm orb}= 0.163$ days \citep{schaefer22a}. 
Thus, it cannot have a large accretion disk like KT Eri.
From such an example, we would stress that only a $V$ flat plateau
does not guarantee the presence of a bright disk during a supersoft
X-ray source phase.

Figure \ref{kt_eri_v339_del_lv_vul_v_small_logscale_no3}d shows the 
$V$ and $y$ light curves both for KT Eri and V339 Del.
The bright plateau
of KT Eri is always above the thick cyan line of $L_V\propto t^{-1.75}$
\citep[universal decline law: ][]{hac06kb}
while the plateau of V339 Del is below this thick cyan line.
This simply means that the shell emission in the $V$ band (typically
[\ion{O}{3}] lines in the nebular phase) is fainter than the bright
large disk in KT Eri.
We can see this in Figure \ref{vflux_o3_he2}a, where the flux of continuum
is larger than that of [\ion{O}{3}] lines. 
\citet{ara13ii} decomposed the doublet [\ion{O}{3}] 4959, 5007\AA\  
lines of KT Eri on day 294 ($=$UT 2010 September 1)
and showed that each component has a broad rectangular shape 
(shocked shell) with an expansion velocity of about $\sim 2,200$ km s$^{-1}$
\citep[second plot from the top in Figure 2 of ][]{ara13ii}. 
As shown in Figure \ref{kt_eri_line_profile}b in Section  
\ref{emission_line_profile_wind} and Figure \ref{kt_eri_line_profile}d
in Section \ref{emission_line_profile_sss}, pedestal (rectangular shape)
emission line profiles are an indication of a spherical shell. 
This confirms our interpretation that the [\ion{O}{3}] lines are originated
from the shocked shell.

In Figure \ref{kt_eri_v339_del_lv_vul_v_small_logscale_no3}d,
the light curves of V339 Del are time-stretched with $f_{\rm s}= 0.76$
and $\Delta V= +0.9$ against those of KT Eri.  
The distance modulus in the $V$ band is obtained to be
\begin{eqnarray}
(m&-&M)_{V, \rm KT~Eri} \cr
&=& ((m - M)_V + \Delta V)_{\rm V339~Del} - 2.5 \log 0.76 \cr
&=& 12.2 + 0.9\pm0.2 + 0.3 = 13.4\pm0.2,
\label{distance_modulus_v_kt_eri_v339_del}
\end{eqnarray}
where we adopt $(m - M)_{V, \rm V339 ~Del}= 12.2$ from \citet{hac24km}. 
Thus, we again obtain $(m-M)_V= 13.4\pm0.2$ for KT~Eri.

Finally, we would like to stress that $y$ magnitude observations are
critically important for novae in order to identify whether or not
a plateau phase is caused by a bright accretion disk. 
However, there are only a few novae that have been well followed
with $y$ magnitude. 
The difference/similarity in the $V,y$ light curves provide 
important information on the nova outburst and binary nature.

\subsection{Time-stretching method in the $B$ and $I_{\rm C}$ bands}
\label{time-streching_b_i_bands}

We apply the time-stretching method to the $B$ and $I_{\rm C}$ 
light curves of KT Eri and obtain the distance moduli both in the $B$
and $I_{\rm C}$ bands.  
Figure \ref{distance_reddening_kt_eri_bvi_xxxxxx}a shows
the $B$ light curves of KT Eri
together with those of LV Vul, V1668 Cyg, and V5584 Sgr.
We apply Equation (\ref{distance_modulus_formula}) in the $B$ band
to Figure \ref{distance_reddening_kt_eri_bvi_xxxxxx}a and obtain
\begin{eqnarray}
(m&-&M)_{B, \rm KT~Eri} \cr
&=& ((m - M)_B + \Delta B)_{\rm LV~Vul} - 2.5 \log 0.36 \cr
&=& 12.45 - 0.1\pm0.2 + 1.1 = 13.45\pm0.2 \cr
&=& ((m - M)_B + \Delta B)_{\rm V1668~Cyg} - 2.5 \log 0.36 \cr
&=& 14.9 -2.45\pm0.2 + 1.1 = 13.45\pm0.2 \cr
&=& ((m - M)_B + \Delta B)_{\rm V5584~Sgr} - 2.5 \log 0.24 \cr
&=& 17.45 -5.5\pm0.2 + 1.55 = 13.5\pm0.2,
\label{distance_modulus_b_v959_mon_lv_vul_v1668_cyg}
\end{eqnarray}
where we adopt the same stretching factor of $f_{\rm s}= 0.36$ for LV Vul
as in Figure \ref{kt_eri_v339_del_lv_vul_v_small_logscale_no3}b,
and $(m-M)_{B, \rm LV~Vul}= 12.45$,
$(m-M)_{B, \rm V1668~Cyg}= 14.9$ and $f_{\rm s}= 0.36$, 
$(m-M)_{B, \rm V5584~Sgr}= 16.7 + 0.75 = 17.45$ and $f_{\rm s}= 0.24$,
all taken from \citet{hac21k}.
Thus, we have $(m-M)_{B, \rm KT~Eri}= 13.47\pm0.2$.

Figure \ref{distance_reddening_kt_eri_bvi_xxxxxx}b
shows the $I_{\rm C}$ light curve of KT Eri together with those of
V5114~Sgr, V1369~Cen, and V496~Sct.
These four $I_{\rm C}$ light curves broadly overlap with each other,
although the $I_{\rm C}$ light curve of KT Eri is located slightly above
the other three light curves in the later phase.
This is because an irradiation effect of a large disk contributes to
the luminosity in the $I_{\rm C}$ band.
Applying Equation (\ref{distance_modulus_formula}) in the $I_{\rm C}$ band
to Figure \ref{distance_reddening_kt_eri_bvi_xxxxxx}b, we obtain
\begin{eqnarray}
(m&-&M)_{I, \rm KT~Eri} \cr
&=& ((m - M)_I + \Delta I_{\rm C})
_{\rm V5114~Sgr} - 2.5 \log 0.48 \cr
&=& 15.56 - 3.1\pm0.2 + 0.8 = 13.26\pm0.2 \cr
&=& ((m - M)_I + \Delta I_{\rm C})
_{\rm V1369~Cen} - 2.5 \log 0.245 \cr
&=& 10.1 + 1.7\pm0.2 + 1.52 = 13.32\pm0.2 \cr
&=& ((m - M)_I + \Delta I_{\rm C})
_{\rm V496~Sct} - 2.5 \log 0.182 \cr
&=& 12.9 - 1.5\pm0.2 + 1.85 = 13.25\pm0.2,
\label{distance_modulus_i_vi_v5114_sgr}
\end{eqnarray}
where we adopt
$(m-M)_{I, \rm V5114~Sgr}= 15.56$ and $f_{\rm s}=0.48$ against that of KT Eri,
$(m-M)_{I, \rm V1369~Cen}=10.1$ and $f_{\rm s}=0.245$, and
$(m-M)_{I, \rm V496~Sct}=12.9$ and $f_{\rm s}=0.182$,
all taken from \citet{hac21k}.
We obtain $(m-M)_{I, \rm KT~Eri}= 13.27\pm0.2$

We plot the distance-reddening relations in these three $B$, $V$,
and $I_{\rm C}$ bands in Figure \ref{distance_reddening_kt_eri_bvi_cross}
by the magenta, blue, and cyan lines, that is, $(m-M)_B= 13.47$,
$(m-M)_V= 13.4$, and $(m-M)_I= 13.27$ together with Equation
(\ref{distance_modulus_v_band}) and
\begin{equation}
(m-M)_B= 5 \log (d / 10{\rm ~pc}) + 4.1 E(B-V), 
\label{distance_modulus_rb}
\end{equation}
and
\begin{equation}
(m-M)_I= 5 \log (d / 10{\rm ~pc}) + 1.5 E(B-V), 
\label{distance_modulus_ri}
\end{equation}
where each coefficient of $E(B-V)$ is taken from \citet{rie85}.
These three lines cross broadly at $d=4.2$~kpc and $E(B-V)=0.08$.
The reddening of $E(B-V)=0.08$ is consistent with that of
\citet{rag09bs}.

\subsection{Color-magnitude diagram and start of nebular phase}
\label{color-magnitude_diagram}

When the $B$ and $V$ light curves of the target nova overlap with
the $B$ and $V$ light curves of the template nova, respectively, with
the same time-stretching method, i.e., with the same time-stretching factor
of $f_{\rm s}$, the intrinsic $(B-V)_0$ color curve of the target nova
also overlaps with the intrinsic $(B-V)_0$ color curve of the template nova.
This implies that the time-stretched $(B-V)_0$-$(M_V-2.5 \log f_{\rm s})$
color-magnitude diagrams of the target and template novae 
overlap with each other \citep{hac19k}. 

Figure \ref{hr_diagram_kt_eri_v339_del_bv_mfs}a shows the 
$(B-V)_0$-$(M_V-2.5 \log f_{\rm s})$ diagram for V339 Del while
Figure \ref{hr_diagram_kt_eri_v339_del_bv_mfs}b is for KT Eri.
In the figures, we adopt $f_{\rm s}= 0.48$ for V339 Del and
$f_{\rm s}= 0.36$ for KT Eri both against LV Vul, and the text of
``$(m-M')_V= 11.4(+0.8)$'' in Figure \ref{hr_diagram_kt_eri_v339_del_bv_mfs}a
means the time-stretching distance modulus in the $V$ band, that is,
$(m-M')_V\equiv (m-M)_V - 2.5 \log f_{\rm s} = 11.4$ and
$(+0.8) \equiv - 2.5 \log f_{\rm s} = -2.5 \log 0.48 = +0.8$ for V339 Del.
Then, the distance modulus in the $V$ band is $(m-M)_V= 11.4 + 0.8 = 12.2$.
The data of V339 Del are taken from \citet{hac24km} while the data
of KT Eri are from SMARTS \citep{wal12bt} and \citet{ima12t}.   
We also plot the time-stretched color-magnitude diagrams for
LV Vul (orange line), V1500 Cyg (green line), and V1974 Cyg (cyan line).
All the data of these novae are the same as those in \citet{hac19k}.

The color-magnitude tracks frequently show splitting into two or three 
(even more) branches depending on the observers (or telescopes).
This is because the response function of each filter is slightly 
different among the telescopes and, as a result, strong emission lines
at/near the blue or red edges of the filters contribute differently
to the $B$ and $V$ band fluxes \citep[see, e.g., Figure 1 of][for
response functions]{mun13dcvf}.

\citet{hac16kb} extensively discussed the color-magnitude diagrams of many
novae and concluded that a turning point/cusp of the track frequently
corresponds to the onset of nebular phase.  In V339 Del, dust shell
formation influenced the color-magnitude track, but its turning point
of the track, from toward blue to toward red,
corresponds to the onset of the nebular phase as shown by the arrow
labeled ``nebular phase''
in Figure \ref{hr_diagram_kt_eri_v339_del_bv_mfs}a, which is detected
with [\ion{O}{3}]/H$\beta > 1$ by \citet{mun13hdc}.
The three tracks (Burlak$+$, SMARTS, and Osaka Kyoiku University(OKU))
of V339 Del show a similar path in the very early phase,
but split into three among the three observatories, and then
merge into almost one path again in the very late phase.
Thus, the nebular phase began at the turning points of these three tracks.

The two tracks of KT Eri (SMARTS and Imamura$+$)
do not show a large split, because there are no stronger emission lines
than the continuum and each emission line cannot contribute significantly
to the $B$ and $V$ band fluxes (as already discussed in Section
\ref{Vy_light_curve_same}).  The bright disk contributes both to 
the continuum flux and permitted emission lines such as H$\alpha$ and
H$\beta$, which cloud the detection of nebular phase.
In the present paper, we use a turning point of $(B-V)_0$ as the
onset of nebular phase (as indicated by the arrow labeled ``nebular phase''
in Figure \ref{hr_diagram_kt_eri_v339_del_bv_mfs}b).
This turning point of $(B-V)_0$
corresponds to day $\sim 40$, as already mentioned in Section   
\ref{Vy_light_curve_same}.  

There is a large difference between V339 Del and KT Eri:
the trend of $(B-V)_0$ color in the later phase of V339 Del
is toward red like the other classical novae (LV Vul, V1500 Cyg, 
and V1974 Cyg), but that of KT Eri seems to stay at/around
$(B-V)_0\sim 0.0$ in the first plateau and $(B-V)_0\sim 0.2$
in the second plateau.  
The color of $(B-V)_0\sim 0.0$ is a typical one for irradiated
accretion disks as frequently observed in recurrent novae like in U Sco
\citep[see Figure 29(d) of ][]{hac21k}. 

We also plot the $(V-I)_0$-$(M_I-2.5 \log f_{\rm s})$ diagram for V339 Del
and KT Eri in Figure \ref{hr_diagram_kt_eri_v339_del_bv_mfs}c and 
\ref{hr_diagram_kt_eri_v339_del_bv_mfs}d,
respectively.  Here, $I$ corresponds to $I_{\rm C}$.  The $(V-I)_0$ color
of KT Eri is close to $0.0$-$0.2$ in the later phase of the outburst
because the disk dominates the optical flux of KT Eri.  The details of
each tracks of the other novae are discussed in \citet{hac21k}.

\end{document}